\documentclass[preprint,12pt,authoryear]{elsarticle}

\usepackage[T1]{fontenc}

\usepackage{amssymb}
\usepackage{amsmath}

\usepackage{graphicx}%
\usepackage{dcolumn}%
\usepackage{bm}%
\usepackage{hyperref}%
\usepackage{caption}
\usepackage{subcaption}
\usepackage{diagbox}
\usepackage{multirow}

\usepackage{xcolor}
\definecolor{cinnamon}{rgb}{0.82, 0.41, 0.12}
\definecolor{pink}{rgb}{0.858, 0.188, 0.478}
\definecolor{black}{rgb}{0.0, 0.0, 0.0}

\usepackage[utf8]{inputenc}
\usepackage[normalem]{ulem}
\graphicspath{{Figures/}}
\usepackage[outdir=./]{epstopdf}
\usepackage{rotating}
\journal{International Journal of Multiphase Flows}
\begin{document}
\begin{frontmatter}

\title{Turbulent pipe flow with spherical particles: drag as a function of particle size and volume fraction}%

\author[l1]{Martin Leskovec}

\author[l1]{Sagar Zade}

\author[l1]{Mehdi Niazi}

\author[l2,l3]{Pedro Costa}

\author[l1,l4]{Fredrik Lundell\corref{cor1}}
\ead{frlu@kth.se}

\author[l1,l5]{Luca Brandt}
\ead{luca.brandt@polito.it}

\cortext[cor1]{Corresponding author}

\affiliation[l1]{organization={FLOW and SeRC, Department of Engineering Mechanics, KTH Royal Institute of Technology},postcode={SE-100 44},city={Stockholm},country={Sweden}}

\affiliation[l2]{
organization={Faculty of Industrial Engineering, Mechanical Engineering and Computer Science, University of Iceland},
addressline={Hjardarhagi 2-6},
postcode={107},
city={Reykjavik},
country={Iceland}}

\affiliation[l3]{
organization={Department of Process \& Energy, Delft University of Technology},
addressline={Leeghwaterstraat 39},
postcode={2628CB},
city={Delft},
country={The Netherlands}}

\affiliation[l4]{
organization={Wallenberg Wood Science Center},
postcode={SE-100 44},
city={Stockholm},
country={Sweden}}

\affiliation[l5]{
organization={Department of Environment, Land and Infrastructure Engineering, Politecnico di Torino},
city={Torino},
country={Italy}}%

\date{\today}%
\begin{abstract}
Suspensions of finite-size solid particles in a turbulent pipe flow are found in many industrial and technical flows. Due to the ample parameter space consisting of particle size, concentration, density and Reynolds number, a complete picture of the particle-fluid interaction is still lacking. Pressure drop predictions are often made using viscosity models only considering the bulk solid volume fraction. For the case of turbulent pipe flow laden with neutrally buoyant spherical particles, we investigate the pressure drop and overall drag (friction factor), fluid velocity and particle distribution in the pipe. We use a combination of experimental (MRV) and numerical (DNS) techniques and a continuum flow model. We find that the particle size and the bulk flow rate influence the mean fluid velocity, velocity fluctuations and the particle distribution in the pipe for low flow rates. However, the effects of the added solid particles diminish as the flow rate increases. We created a master curve for drag change compared to single-phase flow for the particle-laden cases. This curve can be used to achieve more accurate friction factor predictions than the traditional modified viscosity approach that does not account for particle size.
\end{abstract}


\end{frontmatter}

\section{\label{Introduction}Introduction}

It is well known that the presence of a non-negligible mass fraction of a solid phase in a turbulent flow will modulate the flow dynamics in intricate ways \citep{Balachandar2010TurbulentFlow}, especially if the particle size is comparable to or greater than the most minor (Kolmogorov) length scales of the flow \cite{Brandt2021Particle-LadenPerspectives}. This is undoubtedly the case in many industrial applications, where the particle size $d_p$ varies from a few microns to a few centimetres. In contrast, the diameter $D$ of the pipe that conveys the mixture could reach the meter scale \cite{Pullum2018HydraulicStorage}. In addition, it is expected that the flow features a dense concentration of such finite-sized particles, resulting in vibrant dynamics where particle-fluid and particle-particle interactions modulate the flow dynamics \cite{Elghobashi1994OnFlows}. Pressure-driven transport of solid-fluid mixtures in pipelines is a fundamental flow problem of considerable practical significance (e.g., transport of slurries or wastewater, processes related to drinking water treatment, food industry and transport of pulp fibres in paper-making processes). 
In these pipelines, a large portion of the energy is lost due to turbulence-induced drag and a fundamental understanding of how this is modified by the presence of a dispersed solid phase would give clues and insights on how to predict, control and modify the flow, aiming for either lower energy consumption or higher product quality (e.g.\ via increased mixing). Due to the complex interplay between different physical effects and the vast number of input parameters, mainly empirical relationships have been proposed to estimate the pressure drop in hydraulic transport of solid-fluid mixtures \citep[]{DURAND1953BasicResearch, Zandi1967HeterogeneousPipelines}. Over the years, both experiments and simulations have been fraught with high uncertainties, making it challenging to achieve repeatability between two similar tests under the same conditions \citep{Zandi1971AdvancesApplication}. Thanks to the development of efficient algorithms for particle resolved simulations \cite{Uhlmann2005AnFlows,Breugem2012AFlows} and increased measurement capabilities \citep{Brandt2021Particle-LadenPerspectives} turbulent pipe and channel flow laden with mono-dispersed rigid spherical neutrally-buoyant particles have now been studied to a somewhat larger extent with great detail both numerically (see, e.g. \cite{Yu2013NumericalFlow,Zhao2010TurbulenceParticles,Shao2012FullyNumber,Lashgari2014LaminarSuspensions,Picano2015TurbulentSpheres,Lashgari2016ChannelRegime,Fornari2016RheologySuspensions,Costa2016UniversalFlows,NiaziArdekani2017DragSpheroids,Eshghinejadfard2017ImmersedParticles,Peng2019DirectNumber,Yousefi2023OnParticles,Hogendoorn2023FromDNSb}) and experimentally (see e.g. \cite{Zade2018ExperimentalDuct,Leskovec2020PipeTurbulence,Hogendoorn2023FromDNSb}) and thanks to this we have been given insights into the physics governing these flows. 

An approach to modeling suspension flows is to treat the suspension as a continuous medium with an effective viscosity $\nu_e$ compared to the viscosity of pure fluid $\nu$ \citep{Guazzelli2011AUnderstanding}. This has been a common approach in rheology, dating back to the works of \citet{Einstein1911BerichtigungMolekuldimensionen, Batchelor1970TheParticles} and \cite{Batchelor1972TheC2}. These works studied the rheological properties of suspensions using theory and semi-empirical arguments to derive relations for the effective suspension viscosity $\nu_e$ at dilute ($\lesssim 0.01$) and semi-dilute ($\lesssim 0.10$) solid volume fractions. Due to the increasing problem complexity with increasing volume fraction, $\phi$ -- even in laminar flows -- only entirely empirical or semi-empirical fits are available for characterization of the suspension rheology, such as the ones suggested by \citet{Eilers1941DieKonzentration} and \citet{Krieger1959ASpheres}. A drawback of these semi-empirical fits and the continuum approach is that only the solid phase volume fraction is considered, and the size and shape of the suspended particles are left out. The importance of particle size was highlighted by, e.g. \citet{Matas2003TransitionFlow}, who studied how the particle-to-pipe-diameter ratio, $d_p / D $, for mono-dispersed rigid neutrally-buoyant spherical particles, influenced the behaviour of transition to turbulence for particle-laden flows. Non-monotonic trends for the onset of turbulence were found. These experiments were numerically reproduced by \citet{Yu2013NumericalFlow}. More recent experimental work continued along this line \cite{Leskovec2020PipeTurbulence,Hogendoorn2022OnsetFlows} arguing that there is a need to include particle size to predict suspension effects on wall-bounded turbulent flow. 

Using an effective viscosity approach to estimate the turbulent drag was proven by \citet{Costa2016UniversalFlows} to result in a lower drag than what is observed in turbulent suspensions. The discrepancy was explained to originate from the formation of a layer of particles near the solid wall. A scaling law was proposed for predicting the friction Reynolds number as a function of the bulk Reynolds number based on the thickness of the particle wall layer. 
\citet{Lashgari2014LaminarSuspensions,Lashgari2016ChannelRegime} performed numerical simulations of particle-laden channel flow and identified three different flow regimes for suspensions of neutrally buoyant particles: laminar-like regime where viscous stresses dominate, a turbulent-like regime, where Reynolds shear stresses are the main mechanism for momentum transfer and finally, at higher concentrations, a so-called inertial shear-thickening regime where particle stresses are enhanced and are the main contributors to the overall drag. For low Reynolds numbers and low concentrations, viscous stresses dominate, whereas, at higher speeds, the  Reynolds shear stresses take over, and at higher volume fractions, particle stresses are enhanced. Focusing on high Reynolds numbers, a migration of particles to the core region was reported, which was significant in dense cases, $\phi \gtrsim 0.30$. %
\citet{Picano2015TurbulentSpheres} studied dense suspensions in turbulent channel flows up to volume fraction $\phi = 0.20$. They demonstrated that, despite the overall drag continuing to grow, the Reynolds shear stress and velocity fluctuation intensities increased with volume fraction before suddenly declining after a local maximum. The drag increase was attributed to the increased turbulence up to a certain volume fraction threshold and particle-induced stresses. \citet{Yousefi2023OnParticles} investigated dense suspensions (up to $\phi = 0.30$) in turbulent channel flow for Reynolds numbers as high as $10\,000$. They found that the main contributors to overall drag was the particle-induced stresses and the turbulence stresses, only a small part of the drag was attributed to viscous stresses. The flow is turbulent-like for $0<\phi<0.15$ and $3\,000 < Re_b <15\,000$ and moves into the inertial shear-thickening regime when the volume fraction is increased ($0.15<\phi<0.30$) and the flowrate is reduced slightly ($3\,000 < Re_b <10\,000$). Under dense conditions ($\phi =0.30$), the flow becomes turbulent-like again as the flow rate increases and the turbulent stresses dominate the particle-induced stresses. This is attributed to flattening the particle volume fraction profile where fewer particles move to the channel's core. As a result, the particle-induced stress becomes less significant, and the turbulent activity rises. According to their findings, the flow will experience high enough Reynolds stresses and velocity variations to revert to homogeneously turbulent conditions, leaving the particle-induced stresses regime. \citet{Shao2012FullyNumber} studied a turbulent channel flow for two different values $d_p/D$ and volume fraction $\phi \approx 0.07$ and showed that the particle size effect on turbulence is connected to the volume fraction. A higher number of smaller particles are present for a given $\phi$, leading to a larger total particle surface area and, hence, an increase of the Reynolds shear stresses. Similar results were reported by \citet{Eshghinejadfard2017ImmersedParticles}, who found that for two different particle sizes in a turbulent channel flow, the smaller of the two had the most profound effect on velocity statistics. \citet{Peng2019DirectNumber} investigated two different particle sizes in particle-laden turbulent channel and pipe flows and found that smaller particles lead to stronger ejection and sweeping events near the wall due to faster rotation of the small particles and a larger total surface area. They also found that for a given $\phi$ and particle size, there is a stronger migration towards the core for a pipe than a channel flow due to the shrinking volume of the pipe in the centre region. 
Turning our attention to experimental work, \citet{Zade2018ExperimentalDuct} reported similar migration trends for a duct flow compared to a channel flow. They studied three different particle sizes and a range of Reynolds numbers. Also, non-monotonic trends for the pressure drop as a function of particle size were found for high-volume fractions where a reduction of turbulent stresses but an increase of particle-induced stresses was reported. 
The complex interaction of spherical particles and flow structures in a turbulent channel flow was investigated by \citet{VanHout2022CombinedLayer} who found that the particles might affect the evolution of structures in the boundary layer. Only dilute conditions were studied, but the findings indicate that for increased volume fractions the finite-sized particles will modify the near-wall turbulent structures to a large extent and hence impact the wall friction and turbulent drag.
\citet{Hogendoorn2023FromDNSb} performed a combined experimental and numerical study of a turbulent pipe flow of dense suspensions (up to $\phi = 0.47$) where three characteristic cases were identified and a parameter space ($\phi$ vs. $Re$) created. The $Re$ in the study is the suspension Reynolds number based on the effective viscosity from the semi-empirical fit by \citet{Eilers1941DieKonzentration}. The first case was characterised as low $\phi$ and high $Re$ and dominated by turbulence with a homogeneous particle distribution. In the second case, at moderate $\phi$, particle migration towards the pipe centre and forming a solid core were observed at moderate $ \phi$. For even higher $\phi$, the third identified case, the solid core expanded from the centre towards the pipe walls due to maximum packing being reached in the pipe centre. Pressure drop measurements for $\phi$ and $Re$ and non-monotonic trends in the turbulent drag were reported. An increase in drag was seen for the first case, then a decreasing drag for the second case and an even further decrease for the third case. The authors speculated that additional friction from the particle wall layer for low $\phi$ introduces the drag increase first seen, and for higher $\phi$, gradients in the solid volume fractions result in a strong non-uniform effective viscosity in the wall-normal direction. The near-wall region, with its relatively lower $\phi$, acts as a lubrication layer, reducing drag. %

Over the years, progress in understanding particle-laden turbulent transport has been driven by the evolution of experimental and numerical techniques. As stated in the recent review of \citet{Brandt2021Particle-LadenPerspectives}, there is great potential when both approaches are used in a complementary way, as in the studies by \citet{Wang2019InertialComparison} and \citet{Hogendoorn2023FromDNSb}. This is also the approach taken in the present study, where we aim to investigate the effect of neutrally buoyant spherical particles in turbulent pipe flow using interface-resolved direct numerical simulations (DNS) and experimental measurements of pressure drop and velocity statistics from phase-contrast magnetic resonance velocimetry (MRV). Different flow cases are investigated by varying the particle size, concentration, and flow rate of the mixture. Changing the flow rate in the experimental rig is a quick way to vary one of the parameters in the large parameter space. From this parameter variation, exciting cases are identified and further examined using DNS, aiming at revealing the secrets of turbulent particle-laden flow. %
In particular, we examine how the overall pressure gradient changes for the single-phase flow upon the addition of particles, how the velocity field changes, and how these changes are connected to particle migration. We also report a universal drag increase curve based on the scaling of pressure drop measurements for a wide range of flow rates, solid volume fractions and particle sizes.

\section{\label{Methods}Methods}

\subsection{\label{Experimental set-up}Experimental set-up}

\begin{table}[]
\caption{Experimental set-up}
\begin{tabular}{lll}
\textbf{Parameter} & \textbf{Symbol} & \textbf{Value}           \\
\hline
Particle size & $d_p$     & $\{0.5, 1, 2, 3\}$ $mm$ \\
Particle density & $\rho_p$ & $1.03$ $g/cm^3$ \\
Pipe size & $D_{pipe}$  & $\{21, 34, 42\}$ $mm$   \\
Fluid Brix &   $^{\circ}Bx$ & $7.4$ \\
Fluid density & $\rho_f$ & $1.03$ $g/cm^3$\\
Fluid dynamic viscosity & $\mu_f$ & $1.25$ $mPas$\\

\end{tabular}
\label{tab:exp_par}
\end{table}

\begin{figure}[tbp]
\centering
\includegraphics[width=1\linewidth,trim={0 5cm 0 1cm},clip]{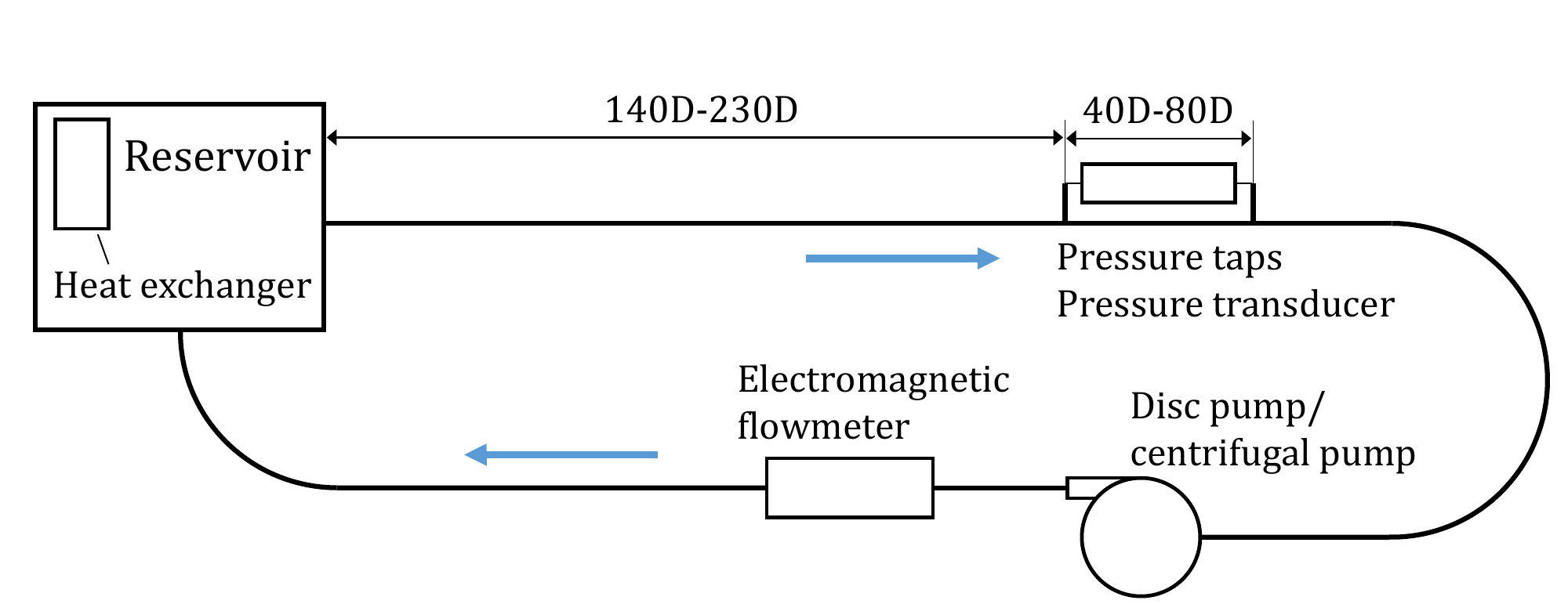}
\caption{\label{fig:rig}Set-up.}
\end{figure}

The experimental setup used to measure the drag of suspension flows is sketched in Figure~\ref{fig:rig}. Three pipes with different diameters ($D = \{21,34,42\}$ $mm$) have been used together with four different particle sizes ($d_p = \{0.5,1,2,3\}$ $mm$) to obtain particle-to-pipe size ratios of $d_p/D = {0.012 - 0.14}$. An aqueous sugar solution, with Brix (g of sucrose per 100g of  solution) $^{\circ}Bx = 7.4\pm0.2$, was used as the suspension solution to match the density of the fluid $\rho_f$ with that of the particles $\rho_p$. The particles were made of STMMA (Styrene-Methyl Methacrylate Copolymer). The addition of sugar increased the dynamic viscosity of the suspension fluid to $\mu_f = 1.25$ mPas, which was measured using a viscometer (Brookfield DVII+Pro, AMETEK Brookfield, Middleboro, MA, USA). The suspension temperature was maintained at 20$\pm$0.5$^\circ$C using an immersion-coil heat exchanger placed inside the conical tank; see Figure~\ref{fig:rig}. A gentle disc pump (Discflo Corporations, CA, USA) was used for the small and large pipe facility, and a centrifugal pump (Flygt model 3085.183, Xylem Water Solutions AB, Sweden) with a modified impeller and volute to allow particles to pass through was used for the medium pipe facility. The pressure drop was measured using differential pressure transducers (Fuji Electric France, S.A.S.) after a development length of at least $140D$ for the three pipes. The two pressure taps were separated by $L=0.8m$. Electromagnetic flowmeters (Krohne Optiflux 1000, Krohne Messtechnik GmbH, Germany) were used to measure the volumetric flow rate $Q$ of the suspension, from which the bulk velocity was calculated as $u_b = Q/A$ with $A=\pi \frac{D^2}{4}$. Drag, expressed as the Fanning friction factor $f$, was calculated from 
\begin{equation}
f = \frac{\tau_w}{\rho_f \frac{u_b^2}{2}} = \frac{\Delta P}{L}\frac{D}{2\rho_f u_b^2}\mathrm{,}
\end{equation}
where $\tau_w$ is the wall shear stress and $\Delta P/L$ the mean streamwise pressure gradient.

In addition to the pressure measurements, we have used Phase-Contrast Magnetic Resonance Velocimetry (PC-MRV or MRV) to obtain the carrier fluid mean velocity and estimates of the variance of the fluid velocity. In this study, MRV allows full-field measurements in fully opaque suspensions, an advantage over traditional experimental methods such as Particle Image Velocimetry (PIV) and Laser Doppler Velocimetry (LDV), which require transparent liquids and solids. We have used a 1~Tesla Magnetic Resonance system (Aspect Imaging, Israel); details of the measurement technique and the MR system can be found in \citet{Dyverfeldt2006QuantificationMRI,Elkins2007MagneticMotion} and \citet{MacKenzie2017TurbulentSlices}. 

\subsection{\label{Numerical set-up}Numerical set-up}

We performed interface-resolved direct numerical simulations (DNS) of turbulent pipe flow laden with finite-sized spheres. The fluid flow is solved on a background regular Cartesian grid with equal spacing $\Delta l$ along all directions, with a volume-penalization immersed-boundary method to describe the pipe geometry (see, e.g. \citet{Kajishima2001TurbulenceShedding,Ardekani2018NumericalParticles}). The immersed-boundary method of \citet{Breugem2012AFlows} (see also \citet{Uhlmann2005AnFlows}) is used to solve the flow around the neutrally buoyant spheres. Short-range hydrodynamic interactions/solid-solid contacts between particles and particles and walls are accounted for using the collision model by \citet{Costa2015CollisionParticles}. The solid–solid contact between the rigid particles is governed by a normal dry coefficient of restitution,
absence of lubrication effects $e_{n,d} = 0.97$, a tangential coefficient of restitution $e_{n,t} = 0.15$, and coefficient of sliding friction $\mu_s=0.2$. Lubrication corrections for the normal lubrication force when the distance between two solid surfaces is small are used as described in \citet{Costa2015CollisionParticles}, with the same model parameters as those used in \citet{Costa2018EffectsSuspensions}.

The general method has been used successfully and validated in previous studies of wall-bounded transport of finite-sized particles (e.g. \citet{Picano2015TurbulentSpheres,Lashgari2016ChannelRegime}.

Turbulent flow is simulated in a domain with periodic streamwise ($x$) direction and no-slip no-penetration along the other directions (e.g. in the absence of volume penalization, a square duct). IBM's volume penalization ensures the solution over a cylindrical pipe with radius $R$ in a flow driven by a uniform pressure gradient that ensures a constant bulk velocity. While the resolution requirements for accurately performing a pipe flow DNS using a volume-penalization IBM are high, resolving the spherical particles with $O(10)$ grid points over the diameter in this fixed-grid solver is what dictates the grid resolution.

Table~\ref{tab:comput_params} reports the physical and computational parameters.

\begin{table}[tbp]
\caption{\label{tab:comput_params}Physical and computational parameters for the interface-resolved particle-laden pipe flow DNS.}

\begin{tabular}{lll}
\textbf{Parameter} & \textbf{Symbol} & \textbf{Value} \\
\hline
Reynolds number & $Re_b$ & $10\,150$\\
Volume fraction & $\phi$ & $\{0, 0.05, 0.20\}$ \\
Pipe-to-particle diameter ratio & $d_p/D$ & $\{0.047, 0.098\}$ \\
Particle-to-element size ratio & $d_p/\Delta \ell$ & $24$\\
\end{tabular}
\end{table}

\subsection{\label{Two-layer model}Upscaled continuum model}

We will use the mean flow model of \citet{Costa2016UniversalFlows} to estimate the overall flow drag under regimes that fall within the model assumptions. The model was developed for a turbulent plane channel flow and was seen to be able to reproduce measurements accurately from direct, particle-resolved simulation data. The present work allows us to further assess the model fidelity against a more comprehensive parameter range. Since we consider a circular pipe here, we summarize the model adapted to this geometry below. Moreover, differently from \citet{Costa2016UniversalFlows}, we derive the model equation for the overall drag expressed in terms of the skin friction coefficient (or Fanning friction factor) $f = \tau_w/(\rho \frac{u_b^2}{2})$, instead of a friction Reynolds number $Re_{\tau}=u_{\tau}h/\nu$ where $u_{\tau}=\sqrt{\tau_w/\rho}$.

The model assumes that the flow domain can be divided into two regions. The first, away from the wall, is the \emph{homogeneous suspension region}, where the particle concentration is approximately uniform, and the apparent particle-to-fluid slip velocity -- the difference between the mean particle and mean fluid velocity -- is negligible. Under these conditions, the flow in this region is assumed to be realistically modeled as a thickened Newtonian fluid. The second region, near the wall, is the \emph{particle-wall layer}, where the confining effect of the wall makes the assumption of a homogeneous region unrealistic. Here, particles tend to skim along the wall with a velocity higher than the local mean fluid velocity.

We therefore assume a two-fluid channel flow with a higher viscosity $\nu_e$ in the homogeneous suspension region, decreasing from $\nu_e$ to $\nu$ in the particle-wall layer. The model assumes that the flow in the homogeneous suspension region corresponds to that of a smaller pipe with an effective suspension viscosity $\nu_e(\phi)$ obtained from well-known correlations from the rheology of particle suspensions \cite{Stickel2005FluidSuspensions}, with a virtual wall origin shifted by $\delta_{pw}(\phi,d_p)$, imposing a stress $\tau^*$ at the edge of the particle-wall layer. Since both regions are subjected to the same pressure gradient, the mean wall shear stress $\tau_w$ can be obtained by linearly extrapolating the total shear stress at the edge of the particle-wall layer, $\tau^*$, to the wall. Hence, the effective pipe radius, effective viscosity, and effective wall shear at the virtual wall origin $\delta_{pw}$ are given by:
\begin{align}
R^* &= R \left(1 - \frac{\delta_{pw}}{R}\right)\mathrm{,}\label{eqn:model_r} \\
\nu_e &= \nu \left(1 + \beta\frac{\phi}{1-\phi/\phi_{\max}}\right)^2\mathrm{, and}\label{eqn:model_nu} \\
\tau^* &= \tau_w\left(1 - \frac{\delta_{pw}}{R}\right)\mathrm{,}\label{eqn:model_tau}
\end{align}
with $\beta=5/4$ and $\phi_{\max}=0.65$ used in the Eilers fit (Eq. \ref{eqn:model_nu}) giving the effective viscosity of the suspension. The virtual wall origin $\delta_{pw}$ is given by the following relation \cite{Costa2016UniversalFlows}:
\begin{align}
\delta_{pw} = C_{pw}\,d_p(\phi/\phi_{\max})^{1/3}\mathrm{,}\label{eqn:model_delta}
\end{align}
with $C_{pw} = \mathcal{O}(1)=1.5$. 

The Colebrook-White formula \citep[]{Colebrook1939TurbulentLaws} for a suspension in a smooth pipe is used for the thickened homogeneous suspension region:
\begin{equation}
\frac{1}{{\sqrt{f^*}}} = -2\log_{10}\left(\frac{2.51}{\sqrt{f^*}Re_{b,e}^*}\right)\mathrm{,}
\end{equation}
where $Re_{b,e}^* = u_bR^*/\nu_e$ is the Reynolds number of the homogeneous suspension region. The values pertaining to the whole pipe can then be determined by using the relations above (Eqs.~\eqref{eqn:model_r}--\eqref{eqn:model_delta}), resulting in:
\begin{equation}
\frac{1}{{\sqrt{f}}} =  -2\xi_{pw}^{1/2}\log_{10}\left(\frac{2.51}{\sqrt{f}Re_b}\frac{\chi_e}{\xi_{pw}^{3/2}}\right)\mathrm{,}\label{eqn:overall_suspension_drag}
\end{equation}
with $\chi_e = \nu_e/\nu$ and $\xi_{pw}=1-\delta_{pw}/R$.
Alternatively, following the same procedure, the explicit Haaland correlation for the friction factor of a smooth pipe \citep[]{Haaland1983SimpleFlow}  becomes:
\begin{align}
\frac{1}{{\sqrt{f}}} = -1.8\xi_{pw}^{1/2}\log_{10}\left(\frac{6.9}{Re_b}\frac{\chi_e}{\xi_{pw}}\right).
\end{align}

We should stress again the assumptions used in the model of \citet{Costa2016UniversalFlows}: i) the slip velocity between phases in the bulk of the flow is negligible; ii) the effective Reynolds number $Re_{b,e}^*$ is high enough to assume scale separation; iii) the extent of the particle-wall layer $\delta_{pw}$ ($\propto d_p$) is much smaller than the pipe radius $R$; iv) the particle concentration in the bulk of the channel is assumed to be close to uniform. When these assumptions are satisfied, we will see that the model produces good quantitative predictions of the present experimental data and the simulations in \citet{Costa2016UniversalFlows}. However, these assumptions -- especially the latter two -- are not always satisfied for the cases presented here. In these cases, the model is bound to produce \emph{qualitatively} incorrect predictions and trends.

\section{\label{Results}Results}

\subsection{\label{Results:Summary}Overview of the experimental and numerical configurations}

\begin{table}[]
\caption{Experimental results.}
\begin{tabular}{lll}
\textbf{Parameter} & \textbf{Symbol} & \textbf{Value}           \\
\hline
Relative particle size & $d_p/D$     & $\{0.012, 0.016, 0.023, 0.029,$\\
&&\hspace{2mm}$0.048, 0.063, 0.1, 0.14\}$ \\
Particle volume fraction  & $\phi$ & $\{0.05, 0.10, 0.20\}$ \\
Bulk Reynolds number & $Re_b$ & $\{10\,000 - 41\,000\}$\\

\end{tabular}
\label{tab:exp_res}
\end{table}

We present experimental data in the form of drag for three volume fractions, nine relative particle sizes and nine different flow rates; see Table \ref{tab:exp_res}. We also report velocity statistics and local volume fraction obtained from MRV for $\phi=0.05$ and $0.20$, $d_p/D =0.1$ and $Re_b\approx 16\,500$ and $20\,600$. Numerical data from DNS are also presented, including drag, velocity statistics and local particle volume fraction for $\phi=0.05$ and $0.20$, $d_p/D = 0.047$ and $0.098$ at $Re_b = 10\,150$. 

As reported in earlier studies (e.g. \citet{Zade2019ExperimentalFluids,Hogendoorn2023FromDNSb,Yousefi2023OnParticles}) the drag increases with solid volume fraction $\phi$ for a given $Re_b$, see Figure \ref{fig:fvsreb}. As $Re_b$ increases, the difference between the single-phase and the particle-laden flow drag decreases, with trends suggesting that the drag eventually approaches the single-phase data regardless of particle size and particle volume fraction, for $\phi = 0.05$ a possible \emph{absolute drag reduction} is seen for high $Re_b$. In what follows, we will analyze these observations in more detail. 

   \begin{figure}[tbp]
     \centering
\begin{subfigure}{.4\linewidth}
   \centering
   \includegraphics[width=1\linewidth]{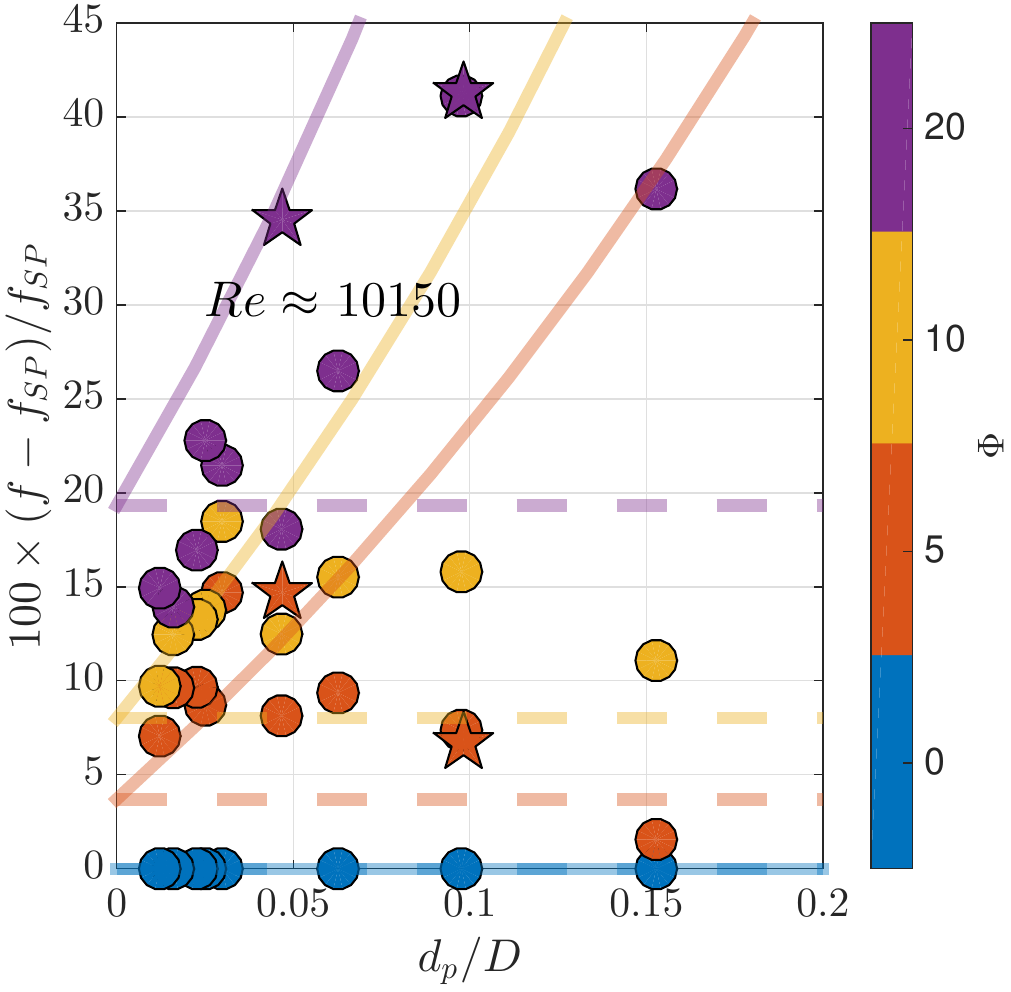}
   \caption{}
\end{subfigure}
   \begin{subfigure}{.4\linewidth}
   \centering
   \includegraphics[width=1\linewidth]{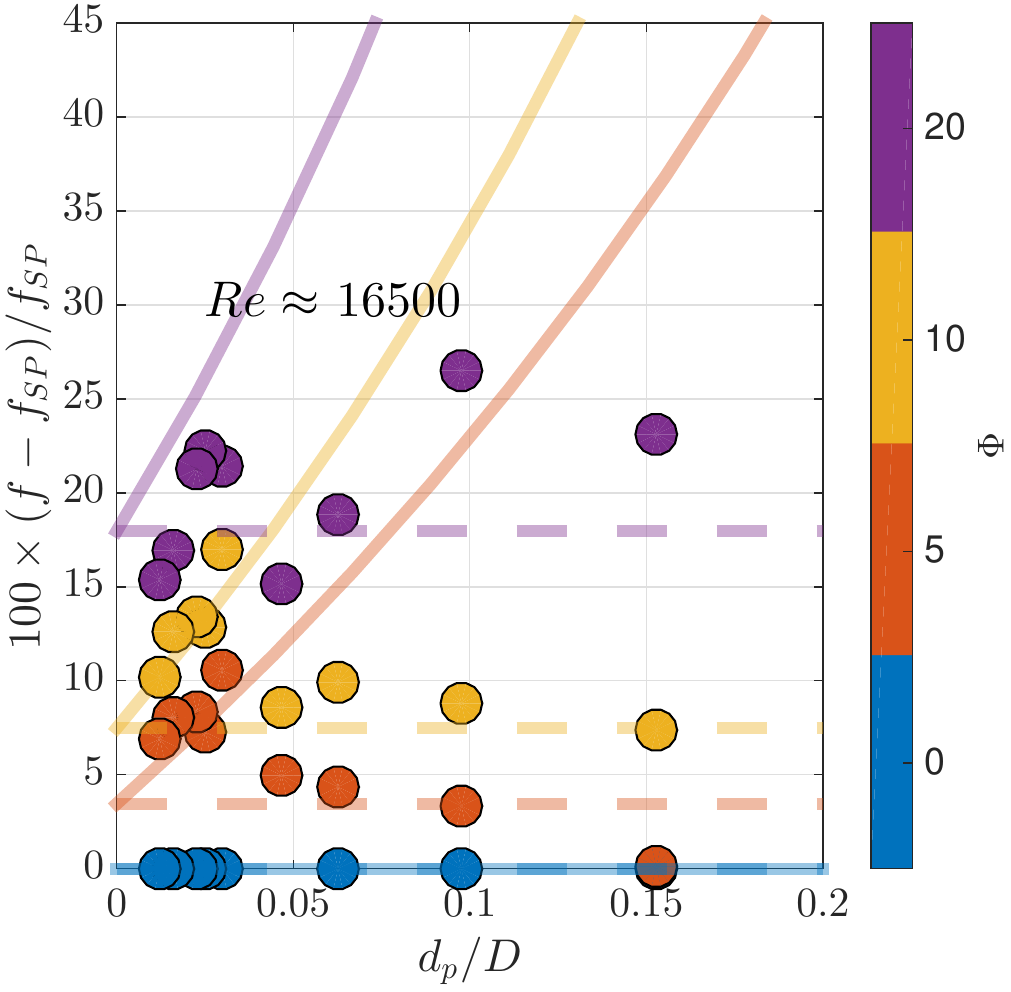}
   \caption{}
     \label{fig:16500}
  \end{subfigure}
  \begin{subfigure}{.4\linewidth}
   \centering
   \includegraphics[width=1\linewidth]{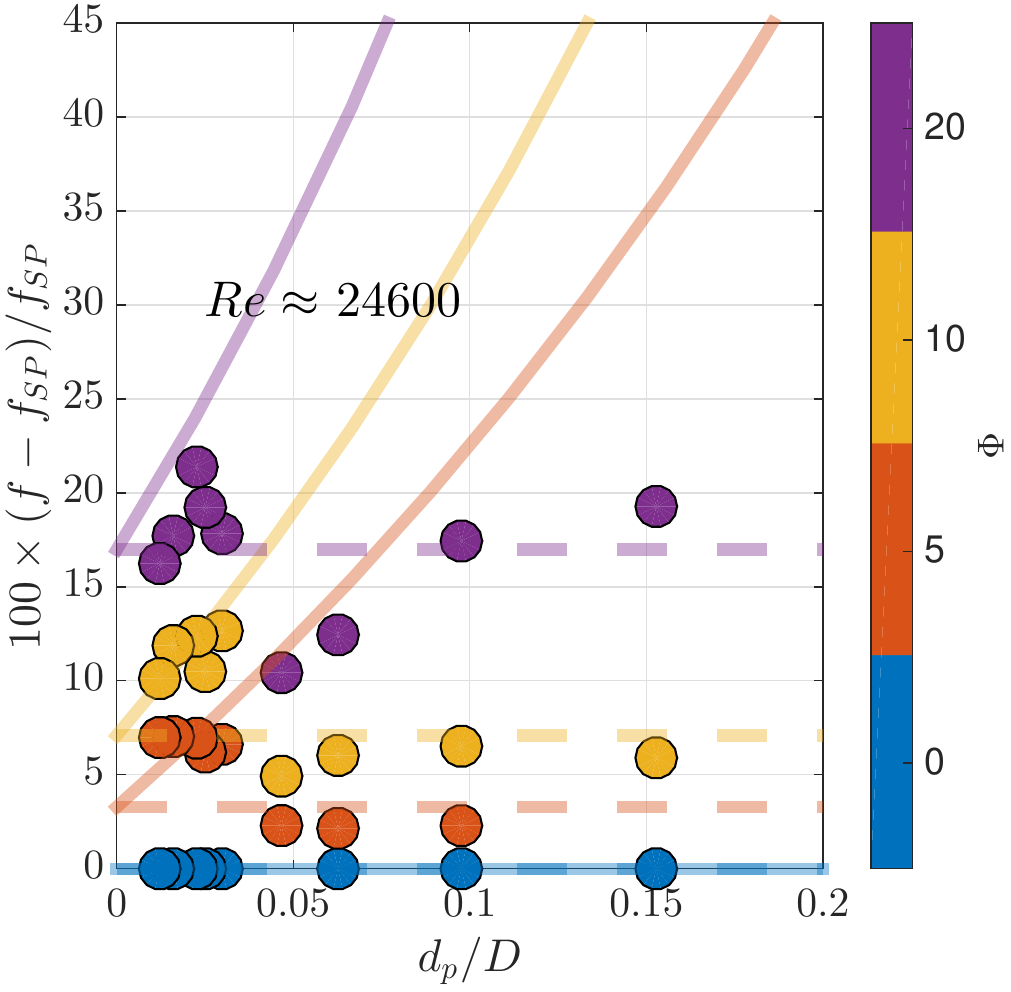}
   \caption{}
   \label{fig:24600}
  \end{subfigure}%
  \begin{subfigure}{.4\linewidth}
   \centering
   \includegraphics[width=1\linewidth]{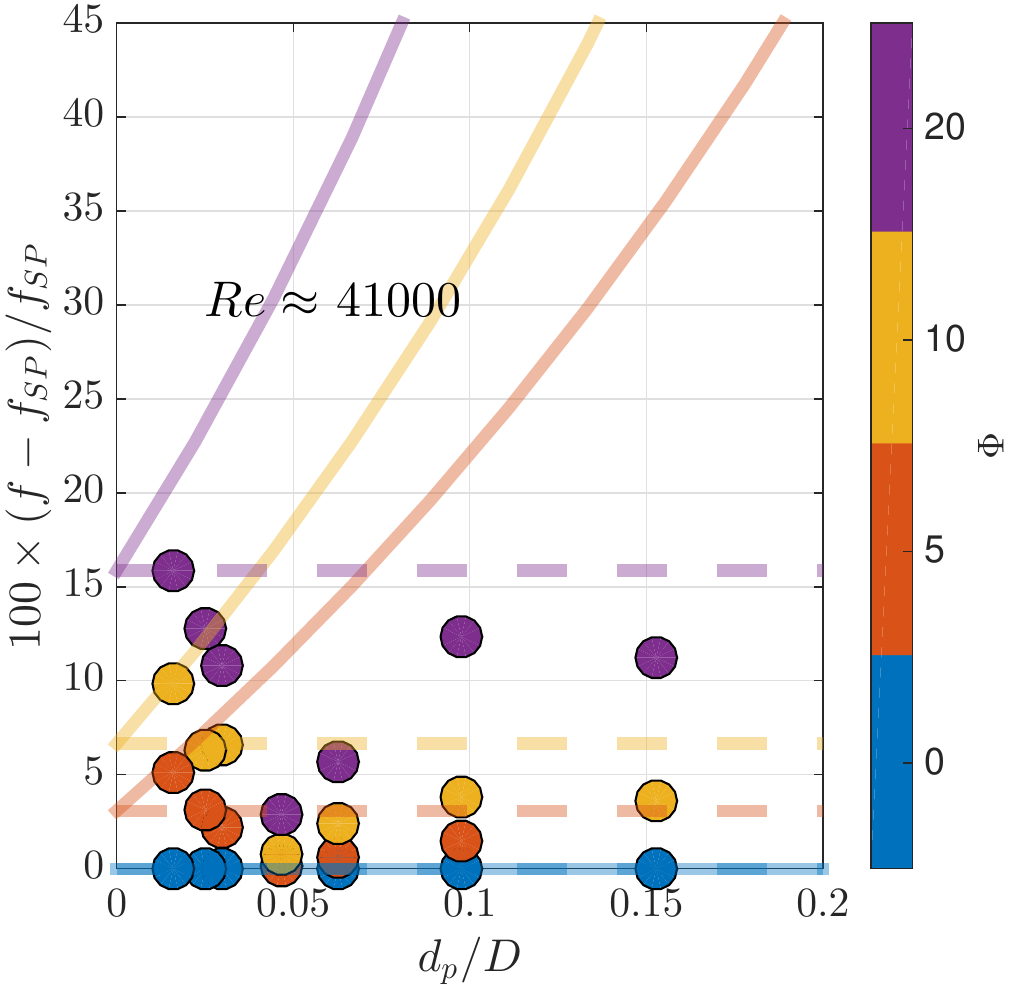}
   \caption{}
   \label{fig:41000}
  \end{subfigure}
  \caption{Relative change in friction factor due to the addition of spheres compared with the single-phase flow of a Newtonian fluid at the same $Re_{b}$. Experiments ($\bullet$), predictions from the particle-wall layer theory \citep{Costa2016UniversalFlows} ($-$) and effective suspension viscosity ($--$). Panel (a) also includes DNS results ($\star$).}
  \label{fig:dD_vs_df}
  \end{figure}

To compare with the numerical data presented here and the continuum model of \citet{Costa2016UniversalFlows}, we break down the trends of pressure drop with particle size for different Reynolds numbers in Figure~\ref{fig:dD_vs_df}. We see the change in friction factor when particles are added to a single-phase flow of a Newtonian fluid as a function of the particle-pipe diameter ratio $d_p/D$. The friction factor predicted by the simplest effective viscosity approach is represented for each $\phi$ as horizontal dashed lines, as this value is independent of the size of the particles. The measured friction factor can be higher and lower than predicted using the suspension effective viscosity above. The solid lines correspond to the friction factor calculated from the particle-wall theory of \citet{Costa2016UniversalFlows}. %

Let us first consider the case at $Re_b \approx 10\,150$ (panel~(a)), for which interface-resolved DNS data is also available. As shown in the figure, the model provides reasonable estimates of the friction factor for $d_p/D \leq 0.03$ and $\phi\leq0.1$. The model predicts an increase in friction factor for larger particle sizes and higher volume fractions, while the measurements and simulations predict a \emph{decrease}. As the Reynolds number increases (panels~(b)-(d)), the friction factor decreases for the experimental measurements, while the model still predicts an increase. These inaccurate model predictions are due to the growing importance of two distinct mechanisms.

First, particles are more prone to migrate towards the core region at large volume fractions due to (inertial) shear-induced particle interactions; see \citet{Fornari2016TheConditions}. The non-uniform particle concentration breaks the validity of the assumption of constant effective viscosity $\nu^e(\phi)$ in the homogeneous suspension region. This mechanism might be modeled by extending the ideas in \citet{Costa2016UniversalFlows} to a variable viscosity profile $\nu^e(\phi)$ in the bulk flow, consistent with the mean local concentration. However, such a modeling approach requires (1) a prediction of the viscosity and particle concentration profiles and (2) a computational strategy for the system using this variable viscosity profile (e.g., a RANS model including a transport equation for the local volume fraction), as in this case there is no simple closed form for the friction factor to derive an equation analogous to Eqn. \eqref{eqn:overall_suspension_drag}.

The second effect causing a deviation from the theory in \citep{Costa2016UniversalFlows} is more difficult to model and occurs at higher flow Reynolds numbers or larger particle sizes. Here, the particle Reynolds number -- a non-dimensional apparent slip velocity -- will increase \cite{Costa2018EffectsSuspensions}. Hence, treating the mean flow in the bulk as a thickened Newtonian fluid becomes unrealistic, irrespective of the particle distribution. This means that, unfortunately, a variable viscosity approach does not suffice for reproducing the physics of the system in this regime, and more complex models accounting for finite-Reynolds number effects are required. A way to model the effect of shear thickening due to non-negligible inertia at the particle scale can be by incorporating the excluded volume effect as described by \citet{Picano2013ShearEffect}. Here, the shadow region of the particle is considered in addition to its physical volume fraction to estimate the effective viscosity of the mixture. The shadow region is some function of the local shear rate. Knowing this along with the local particle volume fraction in the pipe can be a way to extend the analysis of \citet{Costa2016UniversalFlows} to particle-laden pipe flows. Attributing the deviation from the model to the effect of the finite particle Reynolds number is supported by trends in the results seen for increasing Reynolds number in the different panels of Figure~\ref{fig:dD_vs_df}: As the Reynolds number increases, the range of small particle sizes for which the drag increases as predicted by the continuum model becomes smaller and smaller. This trend is particularly evident when comparing the drag-increasing trends for small particle sizes at volume fractions $5$ and $10\%$ in panels (b)-(d) of Figure~\ref{fig:dD_vs_df}. 

For particles of size $d_p/D\approx0.10$, the pressure drop is almost the same in experiments and simulations. For the smaller particles ($d_p/D\approx 0.05$), there is a discrepancy between the results. Despite this discrepancy, we consistently observed in both experiments and simulations that the change in friction factor is a non-monotonic function of the particle size and volume fraction. At low concentrations, the drag decreases as the particle size is increased. At high concentrations, the drag initially becomes higher, but as the particle size is increased, the drag reduces.

\subsection{\label{Results:Observations}Observations and discussion}

\subsubsection{Turbulent drag for $Re_b$ and $Re_e$}

\begin{figure}[tbp]
\centering
\begin{subfigure}{.50\linewidth}
  \centering
  \includegraphics[width=1\linewidth]{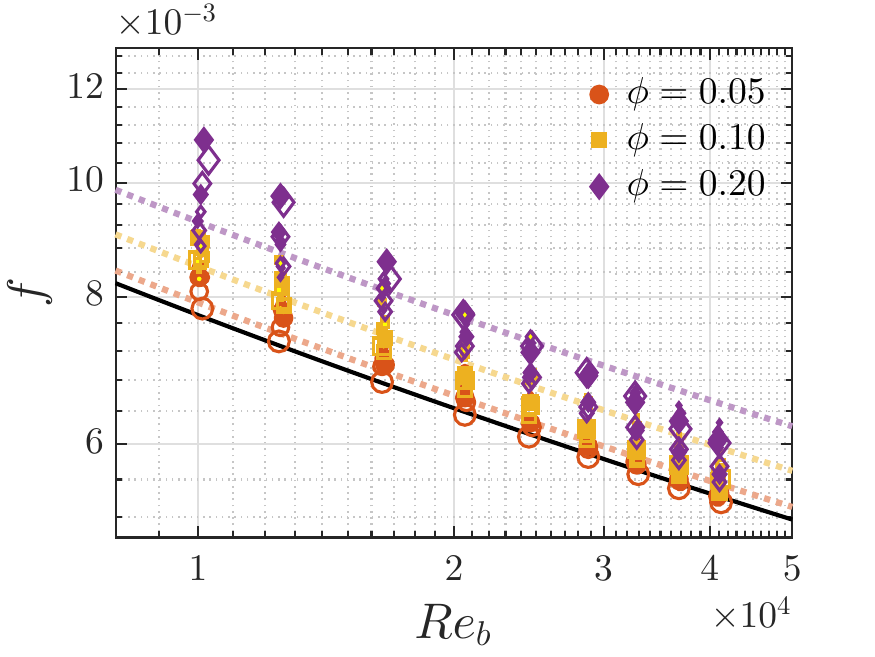}
 \caption{}
  \label{fig:fvsreb}
\end{subfigure}%
\begin{subfigure}{.50\linewidth}
  \centering
  \includegraphics[width=1\linewidth]{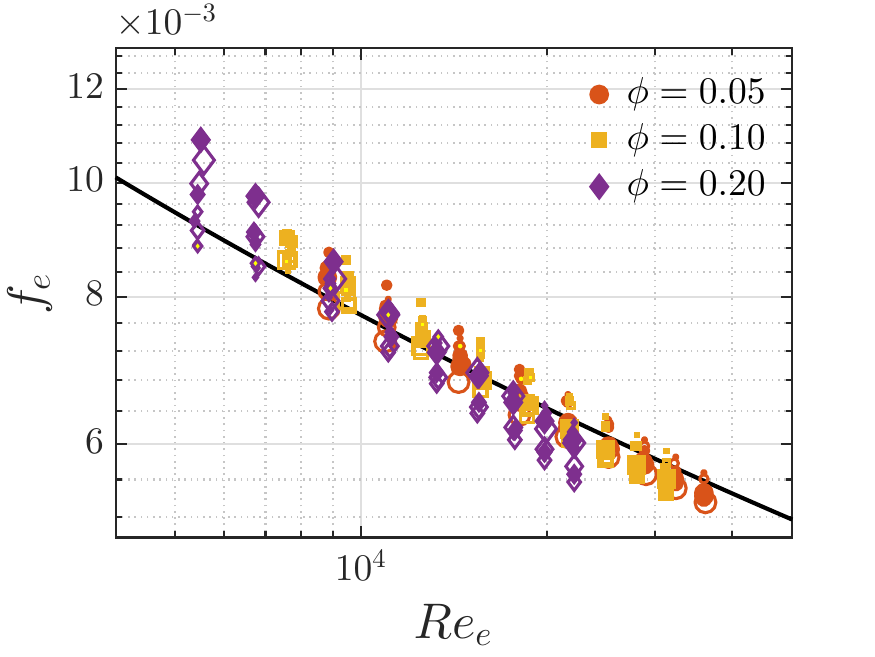}
 \caption{}
  \label{fig:fvsreeff}
\end{subfigure}
\caption{Friction factor vs. Reynolds number for all particle-laden cases compared to single-phase flow represented by the black solid line (correlation from \citet{Duan2012PressureDucts}). (a) Variation of friction factor $f$ with bulk Reynolds number $Re_{b}$. Dashed lines: $2.5\%$, $10\%$ and $20\%$ increase. (b) Same data re-plotted with the friction factor, here renamed $f_e$, expressed as a function of effective bulk Reynolds number, $Re_{e}$, based on an effective viscosity (see Eq. \ref{eqn:model_nu}), due to the addition of solid particles. Filled, unfilled and yellow coloured markers correspond to the three different flow rigs used and the size of the marker corresponds to the relative size of the particles.} 
\label{fig:Re_vs_f}
\end{figure}

As shown in Figure~\ref{fig:fvsreb}, adding the solid phase results in increased drag, here quantified by the Fanning friction factor $f$ as a function of $Re_{b}$. As $Re_{b}$ increases, the deviation of $f$ from single-phase Newtonian fluid, indicated by a black line using the empirical correlation from \citet{Duan2012PressureDucts}, decreases. For reference, the dashed lines in the figure indicate an increase in $f$ of $2.5\%$, $10\%$ and  $20\%$ compared to the single-phase Newtonian case, and the size of the markers increases with the size of the particles (using some representative scale). To elaborate, for the lowest $Re_{b} (\approx 10\,000$) the drag increases from $2\%$ to as much as $42\%$, depending on $d_p$ and $\phi$. For the highest $Re_{b} (\approx 41\,000$), the pressure drop increase is less, from  $-1$ (drag reduction) to $16\%$; a real drag reduction is observed for large $d_p$ and low volume fraction $\phi=0.05$. The particle size $d_p$ seems to play a less significant role than the solid volume fraction, but there is still a dependency on $d_p$ for the friction factor. %
Specifically, at the lowest $Re_{b}$, the largest particles give the highest pressure drop, while at the highest $Re_{b}$, smaller particles result in the highest pressure drop. Note that a similar non-monotonic trend was found in the experiments in a square-duct flow by \citet{Zade2018ExperimentalDuct}.

In Figure~\ref{fig:fvsreeff}, we display the friction factor from the same experiments, now as a function of an effective Reynolds number, with the notation $f_e$. Using the effective viscosity, $\nu_e$, which accounts only for the solid phase volume fraction, yields a reduced Reynolds number $Re_{e} = \frac{u_b D}{\nu_e} $. The effective viscosity is computed using the semi-empirical Eilers fit in Eq. \ref{eqn:model_nu}.
Deviations from the single-phase effective viscosity drag (black line) are especially evident for high volume fractions. At the lower end of the $Re_e$ range all cases except the smallest $d_p$ at highest $\phi_b$ show a drag increase. Differences in $f_e$ appear as the particle size is varied, especially at high $\phi$. From $Re_e\approx10\,000$ and onwards a drag reduction compared to the expected drag for a fluid with an increased effective viscosity is seen for a large share of the studied cases. The deviation from the effective drag is less for small particle sizes, and it is clear that the particle size influences the drag. The study by \citet{Hogendoorn2023FromDNSb} reported a trend of decreasing drag with increasing $Re_e$, and a strong decrease in drag was found for increasing $\phi_b$ (up to $40\%$). However, only one particle size was investigated and as mentioned above, our experiments reveal that $f_e$ has a particle size dependency that cannot be overlooked.

The friction factor $f$ tends towards the single-phase line for all volume fractions when increasing $Re_b$ without considering any modification of the suspension viscosity, see Figure~\ref{fig:fvsreb}. As we will see, this tendency to approach single-phase flow drag and even drag reduction is possibly an effect of particle migration, which forms a layer of ``thinner'' fluid near the wall while having a laminarizing effect on the bulk flow, with mechanisms similar to the turbulent flow of a fluid with a viscosity profile that increases from the wall towards the channel; see \citet{DeAngelis2004DragProfile}. %

As mentioned above, the particle size $d_p$ is a parameter usually not included in the estimation of the effective viscosity. This could explain why a simple normalization of the bulk Reynolds number does not collapse the data well. Note finally that the friction factors $f_e$ for the suspensions of the smallest particles, $d_P/D=0.012$-$0.016$, are very close to the single-phase drag using the effective viscosity for all three values of the volume fraction under investigation. This indicates that models based on the effective viscosity yield good predictions only for specific particle sizes. %

\begin{figure}[tbp]
\centering
\begin{subfigure}{.49\linewidth}
  \centering
  \includegraphics[width=1\linewidth]{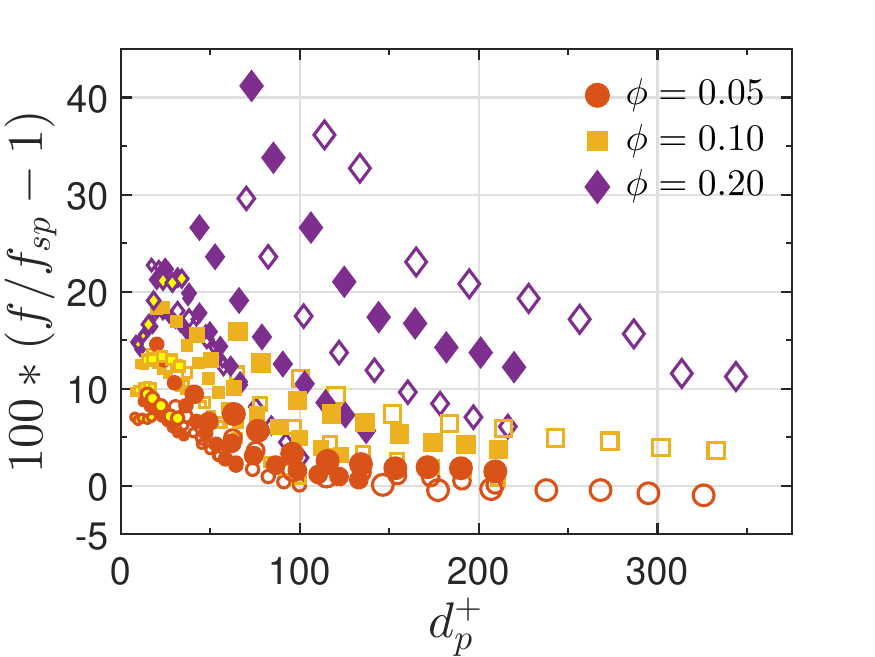}
 \caption{}
  \label{fig:changeinf_dp+}
\end{subfigure}%
\begin{subfigure}{.49\linewidth}
  \centering
  \includegraphics[width=1\linewidth]{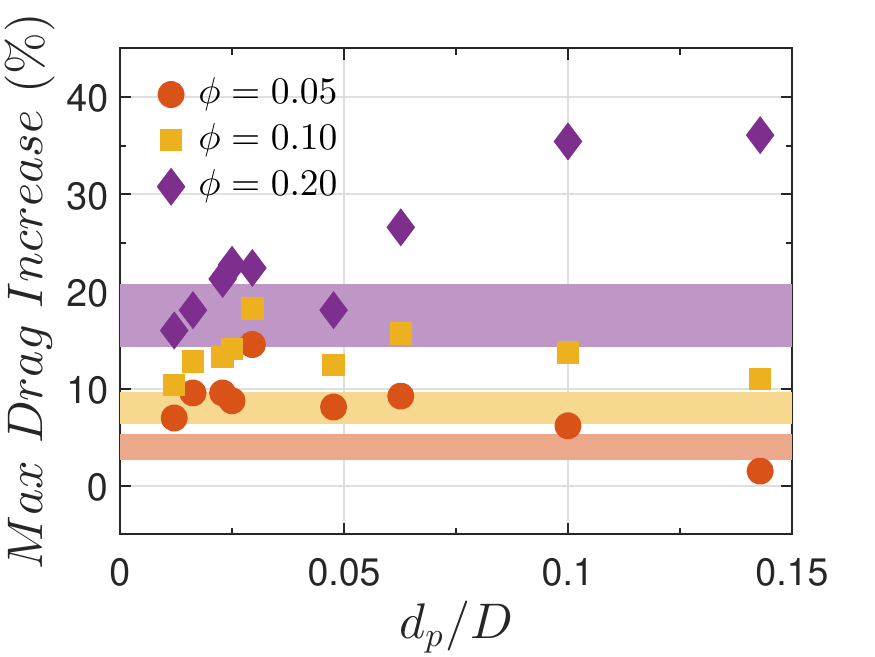}
 \caption{}
  \label{fig:maxdrag_Ddp}
\end{subfigure}
\caption{(a) Maximum increase in friction factor (pressure drop) as a function of $d_p^+$ and (b) maximum increase in friction factor as a function of pipe-to-particle-diameter ratio $D/d_p$. The coloured bands represent the expected drag increase resulting from using an effective viscosity for accounting for added particles. Filled, unfilled and yellow coloured markers in (a) correspond to the three different flow rigs used.}
\label{fig:dp+_maxdrag}
\end{figure}

\begin{figure}
\centering
\begin{subfigure}{.49\linewidth}
  \centering
  \includegraphics[width=1\linewidth]{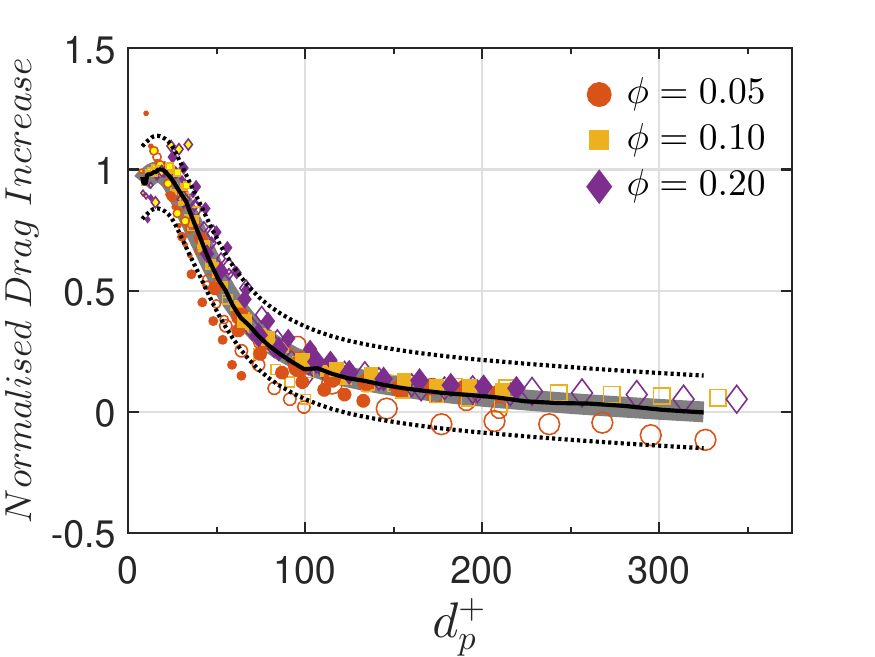}
 \caption{Generalized master curve}
  \label{fig:collapsedcurve}
\end{subfigure}
\begin{subfigure}{.49\linewidth}
  \centering
  \includegraphics[width=1\linewidth]{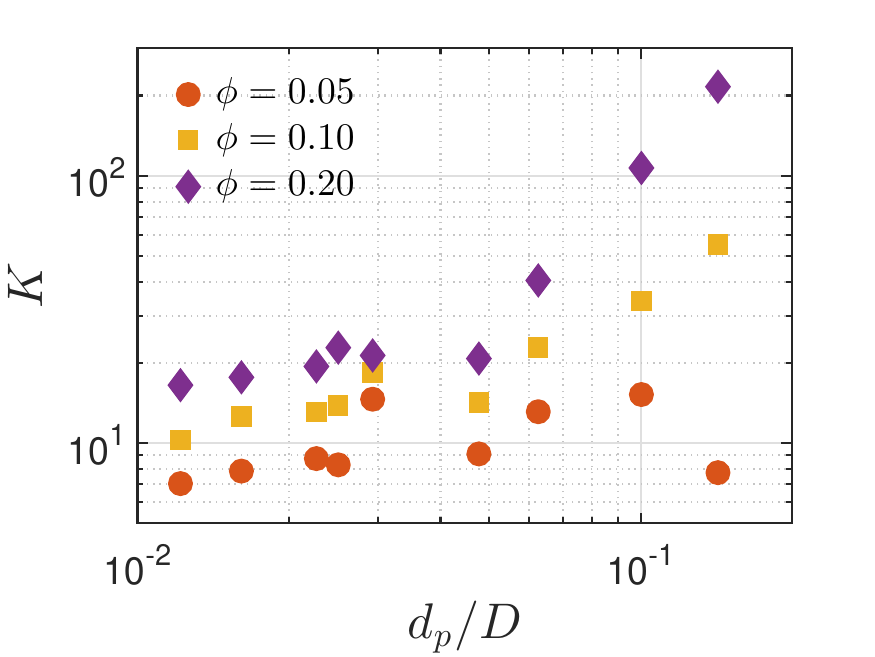}
 \caption{Fitting parameter}
  \label{fig:collapsedcurve_fit_parameter}
\end{subfigure}
\begin{subfigure}{.49\linewidth}
  \centering
  \includegraphics[width=1\linewidth]{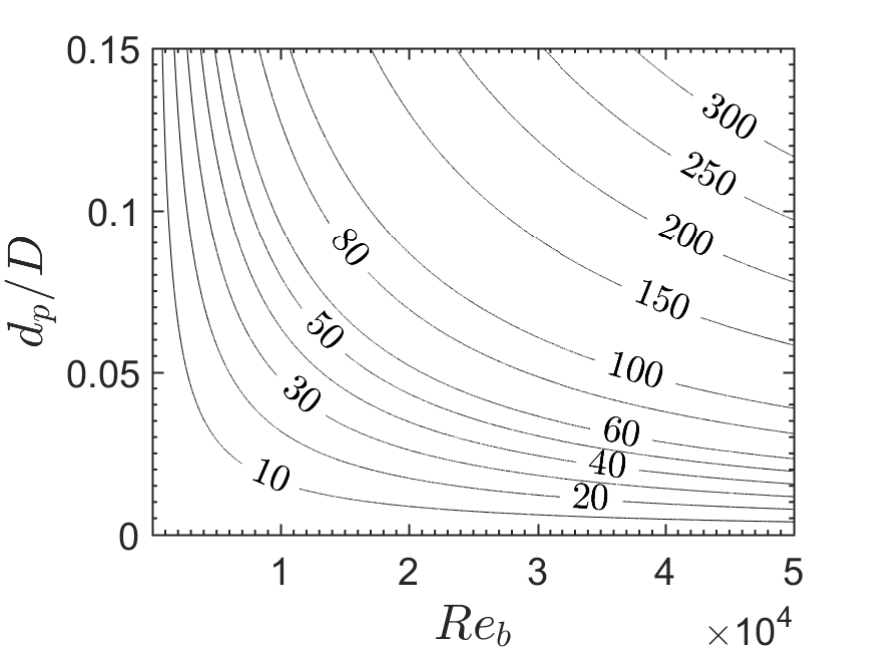}
 \caption{$d_p^+$-space}
  \label{fig:dp+_space}
\end{subfigure}
\caption{The acquired pressure drop data has been collapsed into a generalized master curve, solid black line in (a), for all volume fractions and particle sizes using a fitting parameter for each data set (b). The $d_p^+$-space spanned by $Re_b$ and $d_p/D$ is seen in (c). The information can be used to estimate the pressure drop increase for a turbulent particle-laden pipe flow compared to single phase flow.}
\label{fig:curve_fit}
\end{figure}

\subsubsection{Turbulent drag for $Re_p$ in inner units $d_p^+$}

To further investigate the role of particle size, we show in figure~\ref{fig:changeinf_dp+} the relative change in drag versus the particle Reynolds number in inner units $d_p^+ = d_p u_{\tau} / \nu$. As $d_p^+$ increases, the drag increases and then decreases for $dp/D < 0.016$ for all values of $\phi$. The flow exhibits an absolute drag reduction at the highest $d_p^+$ for the lowest concentration $\phi_b$. The maximum measured drag increase (regardless of $Re_b$) is quantified in Figure~\ref{fig:maxdrag_Ddp}, where it is displayed against $d_p/D$. The trends are similar for the data at 0.05 and 0.10 volume fractions, while there seems to be an abrupt increase for particles larger than $d_p/D > 0.05$ at $\phi=0.20$. 
This increase in maximum drag occurs at low $Re_b$ (cf. Figure~\ref{fig:fvsreb}). Further reduction of $d_p^+$ would decrease the bulk Reynolds number $Re_b$ below $10\,000$, outside of the investigated flow rate range.

The colored bands in Figure~\ref{fig:maxdrag_Ddp} indicate the increase in drag calculated using an effective viscosity that does not depend on the particle size but on $Re_b$. For particle sizes $d_p/D < 0.1$, the maximum drag increase in volume fractions $\phi \leq 0.10$ is underestimated when using a prediction based on the effective viscosity of the suspension. 
At volume fraction $\phi =0.20$, an effective viscosity model can predict the maximum drag increase for some particle sizes $d_p/D \leq 0.05$ and underpredict for others. More than an effective viscosity approach is needed for larger particles to capture the increased drag that comes with added solid particles.

An attempt to collapse all the data in Figure \ref{fig:dp+_maxdrag} is made through an iterative averaging process. Scrutinizing the drag vs $d_p^+$ data plotted in Figure~\ref{fig:changeinf_dp+} it was noted that there appears to be a maximum in the drag increase at $20 < d_p^+ < 25$ for data sets that include this $d_p^+$ and have a $d_p/D > 0.016$. The first step of the averaging process is to obtain an average curve for all data sets that contain this maximum. Index $i$ is used for these datasets. First, $d_{p,c}^+$ (where the index c stands for \emph{collapse}) is selected and normalized functions $g_i(d_p^+)$ are obtained as  $g_i(d_p^+) = \frac{f_i(d_p^+)}{f_i(d_{p,c}^+)}$. The normalization factors $f_i(d_{p,c}^+)$ are denoted $K_i$ and can be taken as an estimate of the maximum drag increase for a given data set. We can now obtain  the average of the normalized drag increase curves $\overline{g}_{init}(d_{p}^+)$ by averaging $g_i(d_p^+)$.

In the second step the data sets that does not include $d_{p,c}^+$ (tehse data sets are indexed with $j$ as $f_j(d_p^+)$) are rescaled with a fitted normalization factor $K_j$. This normalization factor is chosen so that $g_j(d_p^+) = \frac{f_j(d_p^+)}{K_j}$ fits $\overline{g}_{init}(d_{p}^+)$ as well as possible (in a least square sense) in the range of $d_p^+$ where $\overline{g}_{init}$ and the respective data set overlap.

Finally, an ultimate mastercurve $\overline{g}(d_p^+)$ is obtained by averaging all normalized data sets, $f_i(d_p^+)/K_i$ and $f_j(d_p^+)/K_j$, up to $d_p^+=300$. This master curve is shown with the black curve on a gray background in Figure~\ref{fig:collapsedcurve} together with the corresponding fitting parameter $K$ (estimate of the maximum increase in drag in percent) for each data set in Figure~\ref{fig:collapsedcurve_fit_parameter}. The dashed lines correspond to $\pm15\%$. The function values for the generalized master curve together with the fitting parmeters used are tabulated in Tables A.1 and A.2 in appendix \ref{app:appendixA}.%

\begin{table}[h]
\caption{Measured ($f_{measured}$) and estimated ($f_{estimated}$) friction factors compared to single phase flow friction factor ($f_{sp}$). Parameters ($d_p^+$, Normalised Drag Increase (NDI), $K$) computed from test cases in \citep{Hogendoorn2023FromDNSb}.}

  \centering
  \begin{tabular}{lcc}
  
    \textbf{Parameter} & {Case 3} & {Case 4}\\
    \hline
    \hline
    \textbf{$Re_b$}&$7253$ & $9977$\\
    \textbf{$d_p/D$} &$0.058$ &$0.058$ \\ 
    \textbf{$d_p^+$} &$27.4$ & $36.2$\\
    \textbf{$\phi_b$} &$0.089$  & $0.195$ \\
    \textbf{NDI} & $0.93$& $0.80$\\
    \textbf{$K$} & $18.2$& $33.5$\\
    \textbf{$\frac{f_{measured}}{f_{sp}}$} & $18.0$ &$26.8$ \\ \textbf{$\frac{f_{predited}}{f_{sp}}$} &$16.9$ &$16.6$
      \end{tabular}
  \label{tab:test_case}
\end{table}

The information in Figure~\ref{fig:curve_fit} can be used to predict the overall pressure drop for a turbulent flow laden with particles. The particle size ratio ($d_p/D$) and the flowrate ($Re_b$) gives a $dp^+$. The normalized drag increase for a given $dp^+$ and the fitting parameter $K$ for the corresponding particle size $d_p/D$ and volume fraction $\phi$ give the total drag increase (in percentage) compared to a single-phase flow. To test this concept, we use data from \citet{Hogendoorn2023FromDNSb} (Cases 3 and 4 were chosen since the other cases required extrapolation of the $K$ parameter outside the $\phi_b$ range of our data) and estimate the friction factor increase for the partice-laden case compared to the single phase flow, see Table \ref{tab:test_case}. For case 3, we get an estimate of the drag increase of $16.9\%$ where the authors have reported an increase of $18.0\%$. A result that indicates the potential of this generalized master curve. However, using case 4 we get a drag increase estimate of $26.8\%$ compared to the reported $16.6\%$. This case is deemed by the authors to be in transition between two stable particle distribution states, homogeneous distribution and core-peaking distribution. This is a possible explanation for the discrepancy between our estimated drag increase and the measured drag increase. Further experiments at higher volume fractions and more particle sizes would add to a fitting parameter $K$ database to strengthen future estimates.

\subsubsection{Mean velocity and local volume fraction}

 \begin{figure}[tbp]
 \centering
 \begin{subfigure}{.50\linewidth}
  \centering
  \includegraphics[width=1\linewidth]{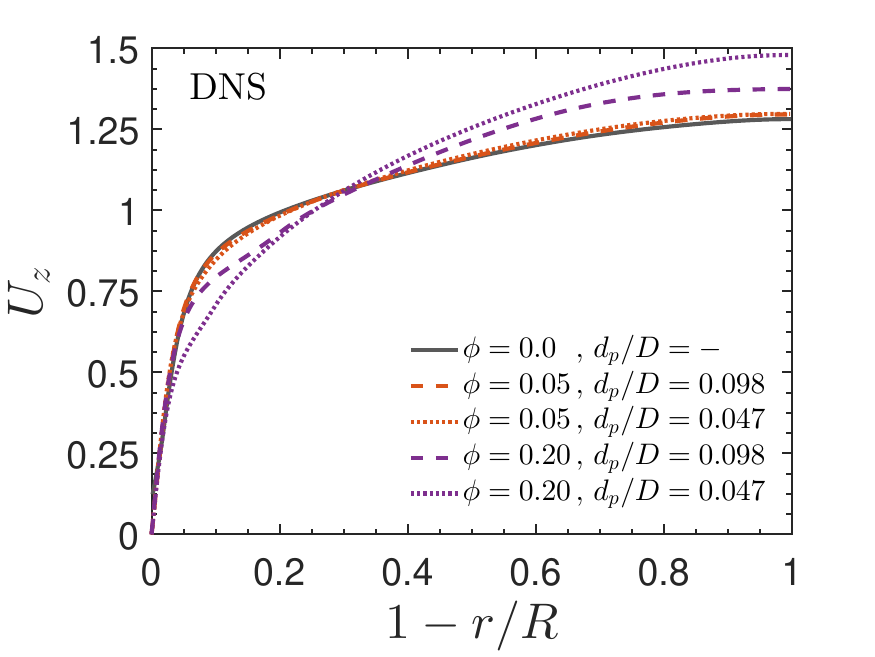}
 \caption{}
  \label{fig:meanvel_DNS}
\end{subfigure}%
\begin{subfigure}{.50\textwidth}
 \centering
  \includegraphics[width=1\linewidth]{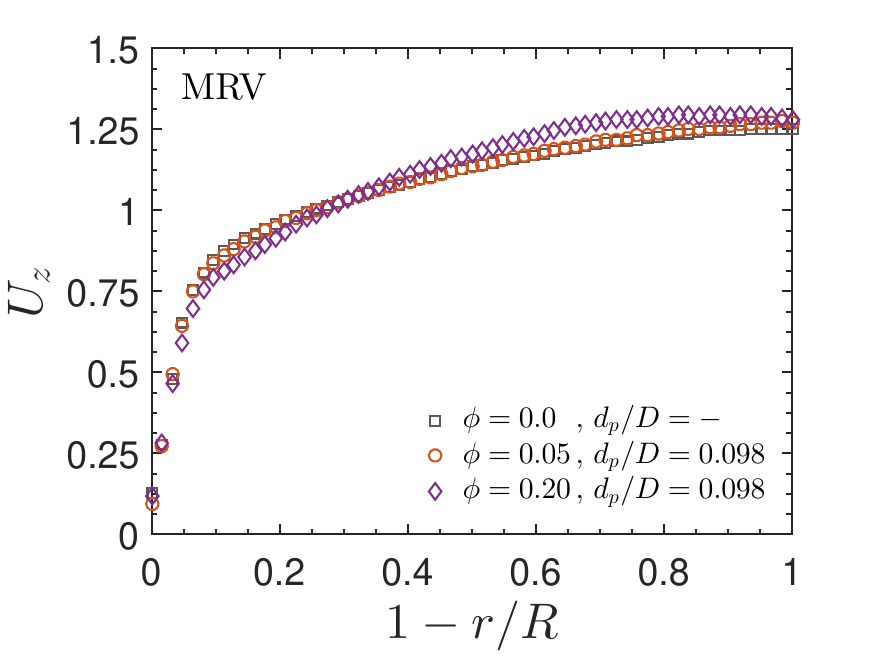}
  \caption{}
  \label{fig:meanvel_EXP16500}
 \end{subfigure}%
 \caption{Profiles of fluid mean velocity from (a) DNS and (b) MRV. Numerical data is acquired at $Re_b = 10\,150$ and the experiment at $Re_b \approx 16\,500$. Small effects are seen for a volume fraction of $0.05$ compared to single-phase flow. }
 \label{fig:meanvel}
 \end{figure}

The mean velocity radial profiles extracted from simulations and experiments are shown in Figure~\ref{fig:meanvel} for different volume fractions and particle sizes. It can be observed that with an increase in the bulk solid volume fraction $\phi_b$, the velocity increases in the central region of the pipe and decreases in the near wall region, resulting in a more parabolic profile resembling that of a laminar flow. The effect is seen in simulations and experiments and is stronger for smaller particles ($d_p/D = 0.098$) at $\phi_b = 0.20$. This is most likely because larger particles accumulate in the center, and the velocity becomes more similar to a plug flow; similar findings have been reported by \citet{Lashgari2016ChannelRegime}. Compared to numerical data, the experimental profiles are obtained at a higher $Re_b$, which could explain the more minor differences between the profiles for different $\phi_b$. The higher $Re_b$ leads to increased turbulent activity, increasing the mixing and distribution of momentum across the pipe. There are no significant differences between the numerical results for the single-phase flow case and the $\phi_b = 0.05$ cases, suggesting that particle size $d_p$ does not reflect significantly in mean velocity profiles at these low concentrations.

\begin{figure}[tbp]
 \centering
 \begin{subfigure}{.49\textwidth}
 \centering
  \includegraphics[width=1\linewidth]{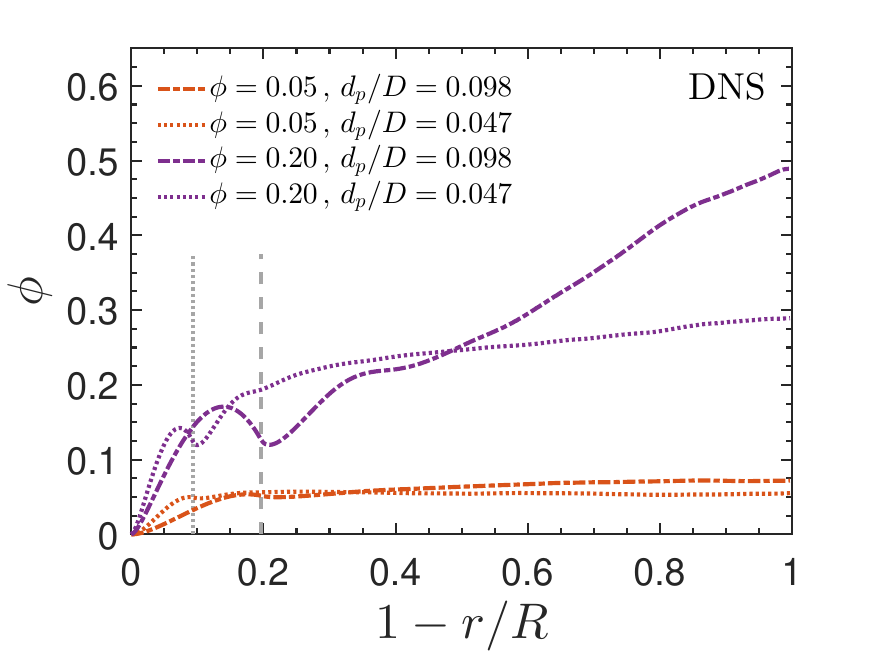}
  \caption{}
  \label{fig:phi_r_DNS}
 \end{subfigure}%
 \begin{subfigure}{.49\textwidth}
 \centering
  \includegraphics[width=1\linewidth]{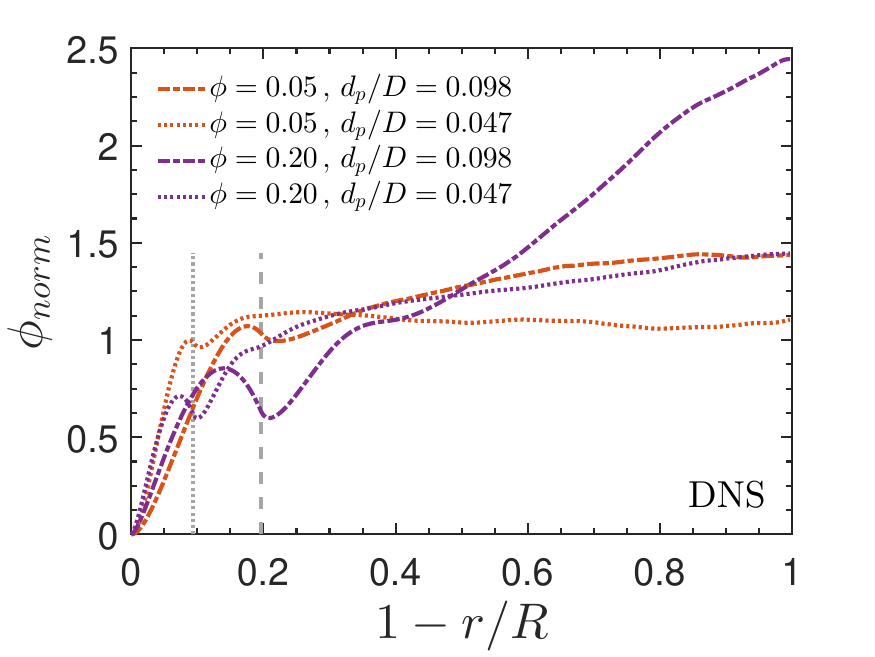}
  \caption{}
  \label{fig:phi_norm_r_DNS}
 \end{subfigure}
 
 \begin{subfigure}{.49\linewidth}
  \centering
  \includegraphics[width=1\linewidth]{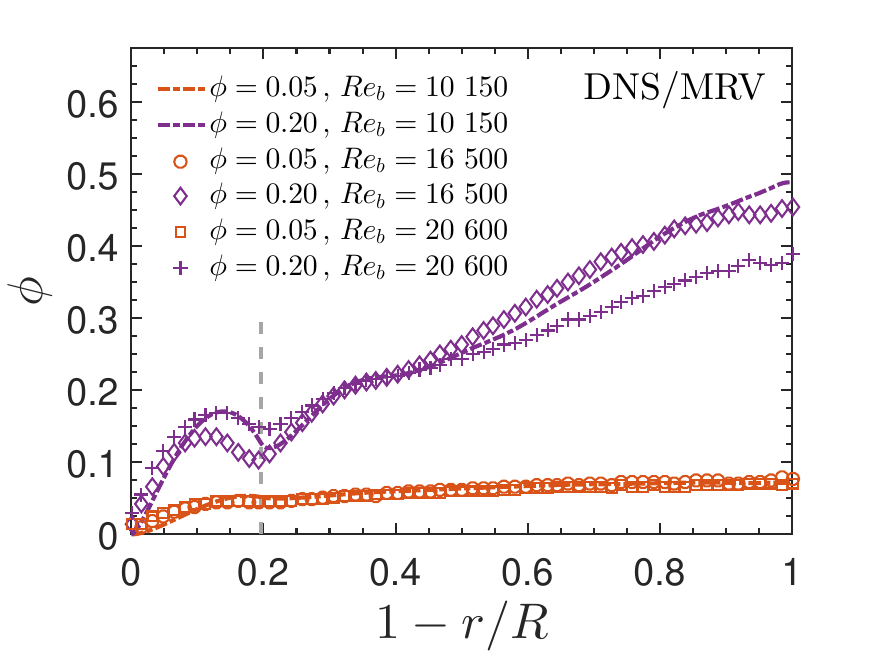}
 \caption{}
  \label{fig:phi_r}
\end{subfigure}%
\begin{subfigure}{.49\textwidth}
 \centering
  \includegraphics[width=1\linewidth]{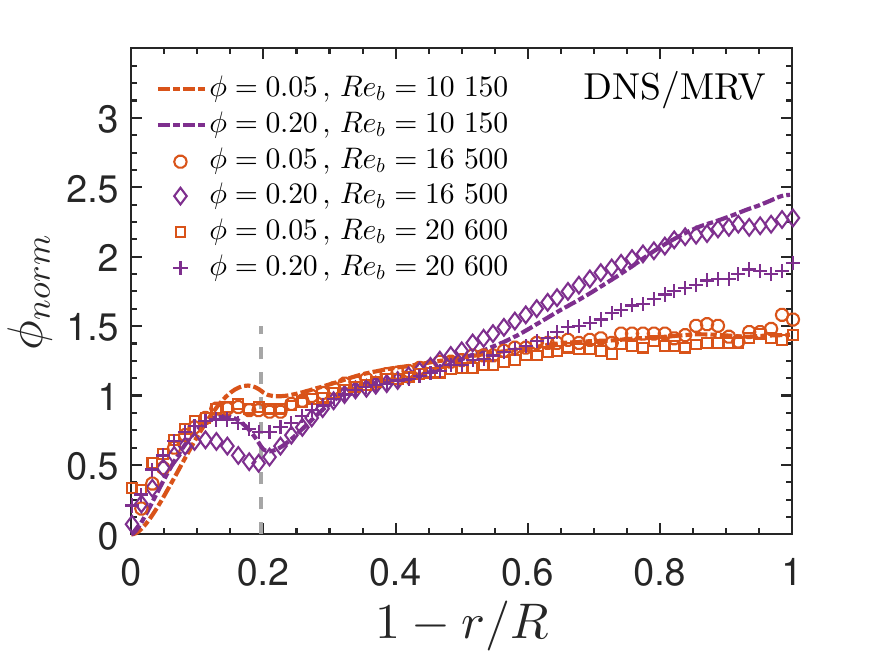}
  \caption{}
  \label{fig:phi_norm_r}
 \end{subfigure}%
\caption{Particle concentration profiles from DNS and MRV of the local volume fraction ($\phi$) and the normalized volume fraction ($\phi_{norm}=\phi/\phi_{tot}$). Grey dashed lines correspond to particle sizes. Peaks in concentration close to the wall show the formation of particle wall layers. (a)-(b) DNS at $Re_b=10\,150$, $\phi$ varies with particle size for dense suspension, constant for semi-dilute. (c)-(d) MRV at $Re_b=16\,500$ (circle, diamond) and $20\,600$ (square, plus) and $d_{p}/D=0.098$. DNS at $Re_b=10\,150$ added for reference.}
  \label{fig:phi_r_DNS_exp}
\end{figure}

The numerical and experimental profiles of the particle concentration are reported in Figure~\ref{fig:phi_r_DNS_exp} for different $Re_b$, $\phi_b$ and $d_p$ to provide insight into the particle migration and its role on the velocity just discussed. 
We start by keeping the flow rate constant and varying the particle size. For low solid volume fractions ($\phi_b=0.05$), the particle size has limited effects on the particle concentration profiles, resulting in an almost homogeneous particle concentration throughout the pipe, see Figure~\ref{fig:phi_r_DNS} and \ref{fig:phi_norm_r_DNS}. The normalized local volume fraction $\phi_{norm}$ is the local volume fraction $\phi$ divided by the bulk solid volume fraction $\phi_b$. For $\phi_b=0.20$, there is an apparent size effect where larger particles migrate towards the centre line, resulting in a local volume fraction of $\phi \approx 0.50$. 
For $\phi_b=0.20$, a distinct peak in the particle concentration is observed close to the wall at a position corresponding to the particle size. This wall peak is associated with the particle-wall layer discussed in \citet{Costa2016UniversalFlows} and is responsible for increased flow drag. A weak dip is seen
for $\phi_b=0.05$ at $1-r/R = d_p/R$, which indicates a weaker wall layer formed already at this volume fraction. 

Now, we keep the particle size constant and change $Re_b$; see Figure~\ref{fig:phi_r} and \ref{fig:phi_norm_r}. There are almost no changes between the investigated flow rates for the low solid volume fraction; the local concentration profiles are very similar. However, for the higher volume fraction, an increase in flow rate decreases the local volume fraction in the pipe centre and increases the concentration in the wall layer. The local particle concentration is still highest in the centre of the pipe, but the migration is less compared to the lower $Re_b$.

\begin{figure}[tbp]
\centering
\begin{subfigure}{.49\linewidth}
\centering
  \includegraphics[width=1\linewidth]{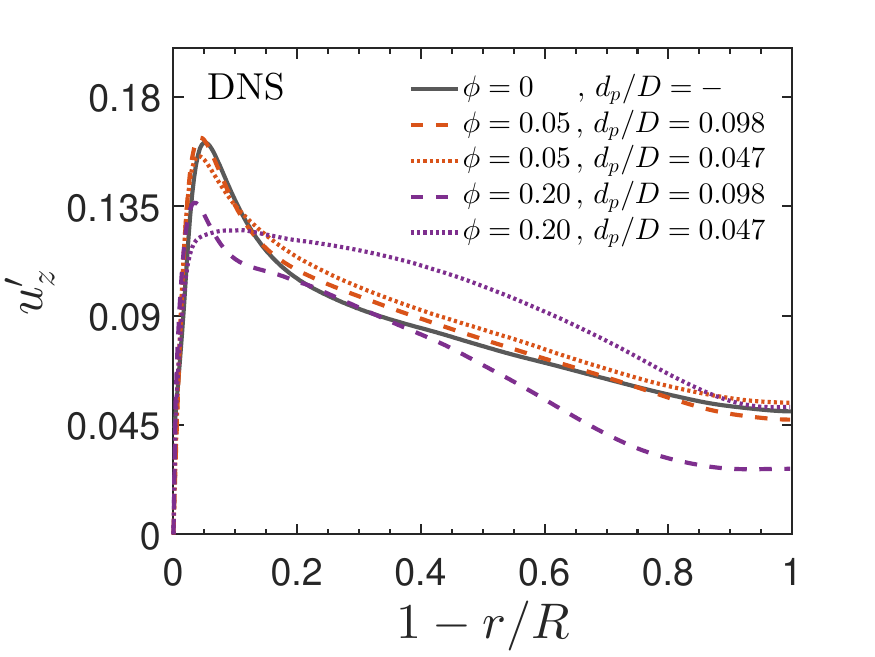}
 \caption{}
  \label{fig:uprimz_DNS}
\end{subfigure}
 \begin{subfigure}{.49\linewidth}
  \centering
  \includegraphics[width=1\linewidth]{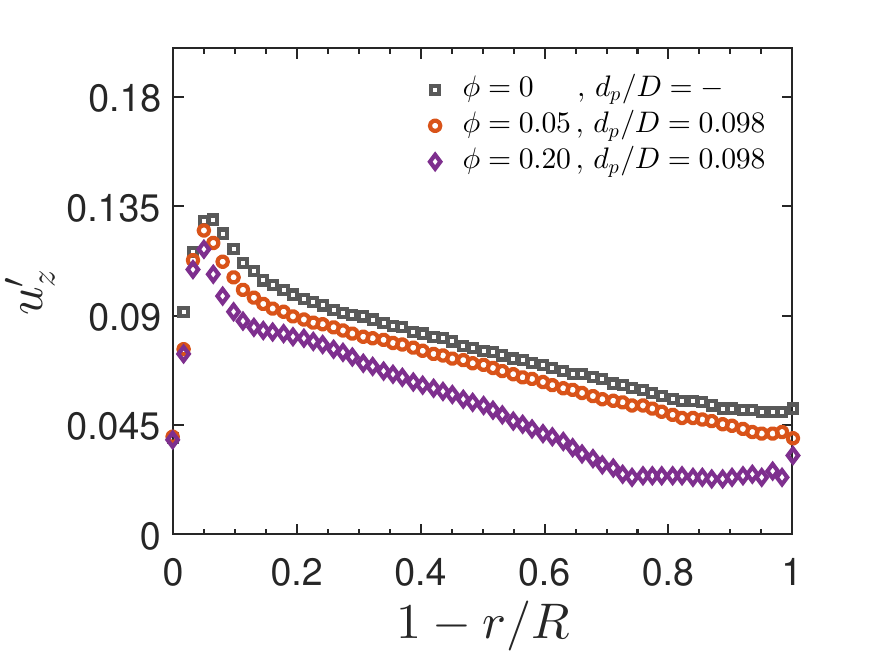}
  \caption{}
  \label{fig:uprimz16500}
 \end{subfigure}
\begin{subfigure}{.49\linewidth}
  \centering
  \includegraphics[width=1\linewidth]{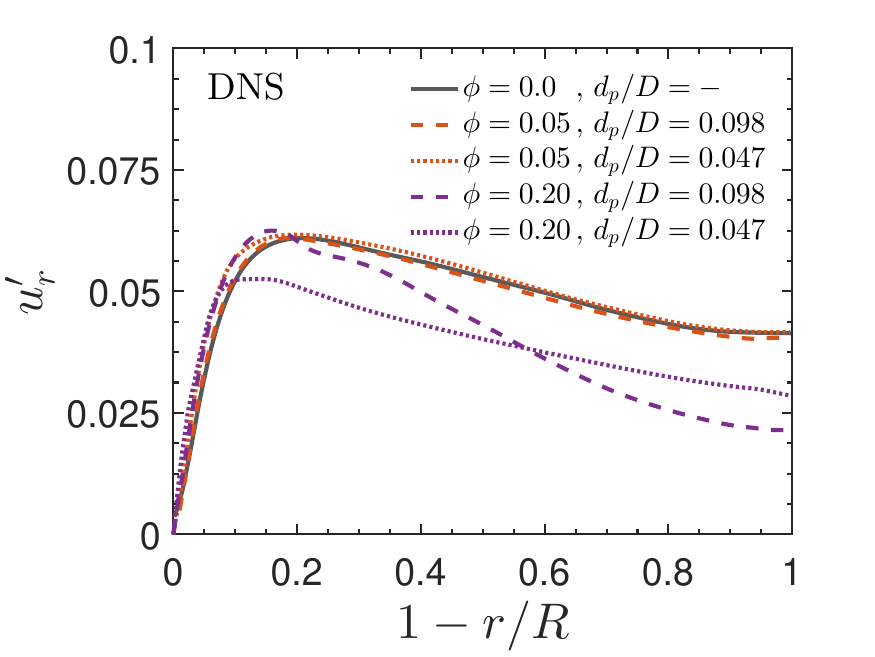}
 \caption{}
   \label{fig:uprimr_DNS}
\end{subfigure}%
\begin{subfigure}{.49\linewidth}
  \centering
  \includegraphics[width=1\linewidth]{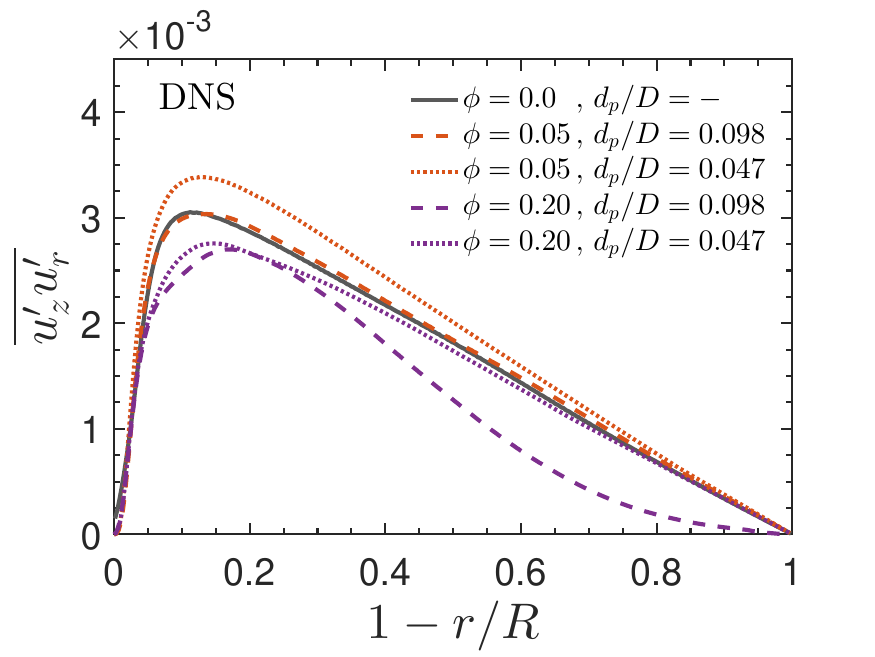}
 \caption{}
  \label{fig:RSS}
\end{subfigure}
\caption{Radial profiles of rms of velocity fluctuations normalized by bulk velocity.
 Streamwise velocity fluctuations from (a) DNS at $Re_{b} = 10\,150$ and (b) from experiments at $Re_{b} \approx 16\,500$. 
 (c) radial velocity fluctuations and (d) Reynolds shear stress from DNS at $Re_{b} = 10\,150$.}
\label{fig:fluctuations}
\end{figure}

We next focus on the fluid velocity fluctuations; see Figure~\ref{fig:fluctuations}. At low bulk concentration, $\phi_b=0.05$, the fluid velocity fluctuations in the streamwise ($u'_z$) and radial ($u'_r$) directions show no significant variations due to particle size; the same pattern as the mean velocity and the local concentration profiles. However, a stronger effect is seen for the Reynolds shear stresses, $\overline{u'_z u'_r}$, for the smaller particles in the near wall region. At $\phi_b=0.20$, the particle size plays a significant role. The near-wall fluctuation peak of $u'_z$ is dampened for both particle sizes. However, compared with $\phi_b=0.05$, the smaller particles enhance streamwise fluctuations in the region between the location of the near-wall peak and the pipe centre. The larger particles result in lower streamwise fluctuations across the entire pipe, with a region in the pipe centre where fluctuations are strongly attenuated. This is related to the increased local solid concentration, seen in Figure~\ref{fig:phi_r_DNS}, due to the 
migration of particles towards the centre. Similar findings have previously been reported for flows at lower Reynolds numbers; see, e.g. \citet{Ardekani2018NumericalParticles}. The dampened fluctuations for the large particles are seen in both DNS and MRV data. The radial velocity fluctuations, $u'_r$, decrease for both particle sizes as $\phi_b$ increases. The Reynolds shear stress profile approaches zero in a region around the pipe centre for large particles at $\phi_b=0.20$, indicating a laminarization of the flow due to the high local volume fraction. In this case, the primary mechanism for momentum transfer is the particle-induced stresses suggesting that this flow case is in the inertial shear-thickening regime as defined by \citet{Lashgari2014LaminarSuspensions}. Comparing DNS and MRV data in Figure~\ref{fig:uprimz_DNS} and ~\ref{fig:uprimz16500} we see that the radial profiles resemble each other for large portions of the pipe. Close to the wall there is however discrepancies were the MRV data has a lower peak in fluctuations compared to the DNS. Estimating fluctuations using the turbulence MRV protocol implemented here comes has some limitations and drawbacks. One such is that in the protocol developed by \citet{Dyverfeldt2006QuantificationMRI} there is an assumption that the velocity distribution is Gaussian. This does not hold close to a solid wall in turbulent flows and it will have implications for the estimated turbulence level there.

\section{\label{Conclusions}Summary and Conclusions}

We have experimentally and numerically studied particle-laden turbulent pipe flow at various flow rates, volume fractions and particle sizes. Our observations are summarized in Table~\ref{tab:summary}. 

For bulk volume fractions $\phi_b=\{0.05,0.10\}$, particles of all sizes have a similar effect on the drag for low Reynolds numbers; there is a minor change in the drag compared to the drag for the unladen case. As the Reynolds number is increased, the drag for medium and large particles approaches the drag of single-phase flow; the small particles, however, act as a pure viscosity enhancer, and the drag for the smaller particles is similar to the drag expected for a single-phase fluid with increased viscosity. Regarding the high volume fraction, drag is higher than single-phase flow for all particle sizes and Reynolds numbers. At higher flow rates, the drag approaches (but does not fully reach) the drag of a single-phase fluid for medium and large particles. The drag for smaller particles is yet again close to the drag of a single-phase fluid with a modified viscosity. This makes us conclude that using a modified viscosity not considering particle size to compute an effective Reynolds number is insufficient to capture the effect of the addition of particles on the friction factor. 

Velocity statistics and local particle concentration have an explicit dependency on particle size. Particle migration towards the core for the larger particles is seen when the volume fraction is high, along with a change in the mean velocity and turbulence statistics compared to the single-phase and low volume fraction cases. The migration towards the centre of the pipe is stronger for the large particles than the small ones. This migration manifests itself in reduced velocity fluctuations and Reynolds stresses in the centre region of the pipe. The smaller particles laminarise the flow, leading to a high centerline velocity, reduction in streamwise velocity fluctuations in the near wall region, and lower velocity fluctuations in the radial direction compared to single-phase flow. Small particles are more evenly distributed in the pipe than large particles, even though a particle-wall layer is seen in both experiments and simulations for both particle sizes. A higher centerline velocity than for single-phase flow is also seen for large particles, but not to the same extent as for small particles. 

\begin{table}[tbp]
\caption{The effect on drag, velocity and particle migration for particle-laden cases compared to single-phase flow. The drag increase is indicated by $\uparrow = 2.5\%$, $\uparrow \uparrow= 10\%$ or $\uparrow \uparrow \uparrow= 20\%$, no increase is indicated by $\approx$. A question mark denotes uncharted territory.}
\resizebox{\textwidth}{!}{%
\begin{tabular}{ll|cc|cc|cc}
\bf{Particle} &  & \multicolumn{2}{c|}{\bf{Low conc.}} & \multicolumn{2}{c|}{\bf{Medium conc.}} & \multicolumn{2}{c}{\bf{High conc.}} \\
\bf{size} & & $Re<20k$ & $Re>30k$ & $Re<20k$ & $Re>30k$ & $Re<20k$ & $Re>30k$ \\ \hline \hline
\multirow{3}{*}{\bf{Small}} & Drag & $\uparrow$ & $\uparrow$ & $\uparrow \uparrow$ & $\uparrow$ & $\uparrow \uparrow$ &   $\uparrow \uparrow$ \\
& Velocity statistics & Limited effect & & & & Laminarisation of $U$,\\
& & & & & & $u'_z$ incr. $0.2$$<$$r/R$$<$$0.8$ & \\
& Particle distribution& Uniform & & & & Wall layer, \\
& & & & & & uniform in center & \\ \hline
\multirow{3}{*}{\bf{Medium}} & Drag &  $\uparrow$ & $\approx$ &   $\uparrow \uparrow$ & $\approx$ &  $\uparrow \uparrow \uparrow$ & $\uparrow$ \\
& Velocity statistics & ? & & & & & \\
& Particle distribution & ? & & & & & \\ \hline
\multirow{3}{*}{\bf{Large}} & Drag &  $\uparrow$ & $\approx$ & $\uparrow$ & $\uparrow$ & $\uparrow \uparrow \uparrow$ & $\uparrow \uparrow$ \\ & Velocity statistics & Limited effect & & & & Increased $U_{cl}$,\\ 
& & & & & &reduced $u'_z$ & \\ & Particle distribution & Uniform & & & & Wall layer,\\
& & & & & & incr. towards center&                    
\label{tab:summary}
\end{tabular}}
\end{table}

The drag increase as a function of particle Reynolds number in inner units shows a similar trend for all particle sizes and concentrations: an initial increase to a local maximum and then a decrease towards the drag of a single-phase turbulent flow. This is in agreement with \citet{Yousefi2023OnParticles}, who predicted that at a high enough Reynolds number, the flow would revert to single-phase conditions due to turbulent stresses overtaking particle-induced stresses. Experiments at higher volume fractions would aid in concluding if this also holds under dense conditions ($\phi_b \geq 0.30$).

Our work shows that simplistic numerical models (like effective viscosity models or continuum models like the one of \citet{Costa2016UniversalFlows}) cannot capture the complexity of turbulent particle-laden flows. We believe that the validity of a continuum no longer holds as I) the local apparent viscosity changes as particles migrate towards the core and II) the particle Reynolds number increases -- causing an apparent slip velocity, and more complex models that account for finite particle Reynolds number effects are required. Utilizing particle volume fraction profiles to achieve a varying apparent viscosity was tried in this work but failed to generate high-accuracy results. Note also that such a modelling approach requires first computing the particle volume fraction profiles, which is challenging. 

For now, a way to estimate the overall drag change compared to single-phase flow is to use the master curve we created and the flow parameters for the case in question. This will give a higher accuracy in the drag prediction than if a traditional modified viscosity approach is used.

\bibliographystyle{elsarticle-harv}
\bibliography{referencesFLU}%

\newpage
\appendix
\section{Generalized master curve}
\label{app:appendixA}
\setcounter{table}{0}

\begin{table}[h]
\caption{Function values for Normalised Drag Increase vs. $d_p^+$.}
\begin{center}
\begin{tabular}{|cc||cc||cc|}
\hline
$d_p^+$      & $NDI$ & $d_p^+$      & $NDI$ & $d_p^+$      & $NDI$      \\
\hline
\hline
8    & 0.970 &28.2   & 0.919  & 99.3 & 0.176 \\
8.62 & 0.945 & 30.4 & 0.894 & 107 & 0.179 \\
9.28 & 0.942 & 32.7 & 0.869  & 115 & 0.156 \\
9.99 & 0.943 & 35.2 & 0.819  & 124 & 0.138 \\
10.8 & 0.979 & 37.9 & 0.769 & 134 & 0.127 \\
11.6 & 0.980 & 40.8 & 0.716 & 144 & 0.112 \\
12.5 & 0.980 & 44.0 & 0.656  & 155 & 0.0975  \\
13.4 & 0.984 & 47.3   & 0.601 & 167 & 0.0866 \\
14.5 & 0.989 & 51.0  & 0.546 & 180 & 0.0760\\
15.6 & 0.989 & 54.9  & 0.502 & 193 & 0.0688  \\
16.8 & 0.999 & 59.1 & 0.438 & 208 & 0.0590 \\
18.1 & 1.000 & 63.7 & 0.388 & 224 & 0.0429 \\
19.5 & 0.999 & 68.6 & 0.354  & 242 & 0.0360 \\
21.0 & 0.990 & 73.9 & 0.311 & 260  & 0.0336 \\
22.6 & 0.978 & 79.5   & 0.273 & 280  & 0.0240 \\
24.3 & 0.964 & 85.6  & 0.239 & 302 & 0.0077 \\
26.2 & 0.943 & 92.2 & 0.207 & 325 & -0.00298\\
\hline

\end{tabular}
\end{center}
\end{table}

\begin{table}[h]
\caption{Fitting parameter $K$ (in percent) for $d/D$ and $\phi$.}
\begin{center}
\begin{tabular}{|l|c|c|c|}

\hline
\textbf{$d/D$}      & \textbf{$\phi=0.05$}     & \textbf{$\phi=0.10$}     & \textbf{$\phi=0.20$}     \\
\hline
0.1429 & 8.30 & 63.0  & 225 \\
0.1      & 6.20 & 15.2 & 43.3 \\
0.0625   & 13.1  & 22.8 & 40.8 \\
0.0476 & 9.11 & 14.1  & 20.5 \\
0.0294 & 14.6 & 18.3 & 21.3 \\
0.025    & 8.26 & 13.9 & 22.7  \\
0.0227 & 8.78 & 13.1 & 19.3 \\
0.0161 & 7.88 & 12.6 & 17.6 \\
0.0122 & 7.04 & 10.2 & 16.4 \\ 
\hline
\end{tabular}
\end{center}
\end{table}

\end{document}